\documentclass[12pt]{iopart}
\usepackage{graphicx}
\usepackage{wrapfig}
\usepackage{setstack}
\usepackage{setspace}
\usepackage{breqn}

\newcommand{\ombar}{\bar{\omega}}
\newcommand{\kbar}{\bar{k}}
\newcommand{\shat}{\hat{s}}
\newcommand{\rbar}{\bar{\rho}}
\newcommand{\dbar}{\bar{\delta}}
\newcommand{\dhat}{\hat{\delta}}
\newcommand{\epbar}{\bar{\epsilon}}
\newcommand{\vbar}{\bar{V}_{\parallel}}
\newcommand{\rlns}{\frac{R}{L_{ns}}}
\newcommand{\rlts}{\frac{R}{L_{Ts}}}
\newcommand{\rlus}{\frac{R}{L_{us}}}

\newcommand{\la}{\left\langle}
\newcommand{\ra}{\right\rangle}

\begin{document}

\title{Tractable flux-driven temperature, density, and rotation profile evolution with the quasilinear gyrokinetic transport model QuaLiKiz}
%\vspace{-2 mm}
\author{J. Citrin$^{1,2}$, C. Bourdelle$^2$, F. J. Casson$^3$, C. Angioni$^4$, N. Bonanomi$^{5,6}$, Y. Camenen$^7$, X. Garbet$^2$, L. Garzotti$^3$, T. G\"{o}rler$^4$, O. G\"{u}rcan$^8$, F. Koechl$^9$, F. Imbeaux$^2$, O. Linder$^{1,10}$, K. van de Plassche$^{1,10}$, P. Strand$^{11}$, G. Szepesi$^{3,6}$ and JET Contributors$^*$}

%\address{JET-EFDA, Culham Science Centre, Abingdon, OX14 3DB, UK}
\address{$^{1}$DIFFER - Dutch Institute for Fundamental Energy Research, De Zaale 20, 5612 AJ Eindhoven, the Netherlands}
\address{$^{2}$CEA, IRFM, F-13108 Saint Paul Lez Durance, France}
\address{$^{3}$CCFE, Culham Science Centre, Abingdon, Oxon, OX14 3DB, UK}
\address{$^{4}$Max Planck Institute for Plasma Physics, Boltzmannstr. 2, 85748 Garching, Germany}
\address{$^{5}$Department of Physics `G. Occhialini', University of Milano-Bicocca, 20126 Milano, Italy}
\address{$^{6}$CNR, Istituto di Fisica del Plasma `P. Caldirola', 20125 Milano, Italy}
\address{$^{7}$CNRS, Aix-Marseille Univ., PIIM UMR7345, Marseille, France}
\address{$^{8}$LPP, Ecole Polytechnique, CNRS, 91128 Palaiseau, France}
\address{$^{9}$\"{O}AW/ATI, Atominstitut, TU Wien, 1020 Vienna, Austria}
\address{$^{10}$Science and Technology of Nuclear Fusion, Department of Applied Physics, Eindhoven University of Technology, PO Box 513, 5600 MB, Eindhoven, The Netherlands}
\address{$^{11}$Department of Earth and Space Sciences, Chalmers University of Technology, SE-412 96 G\"{o}teborg, Sweden}
\address{$^{*}$See the author list of “Overview of the JET results in support to ITER” by X. Litaudon \textit{et al.} to be published in Nuclear Fusion Special issue: overview and summary reports from the 26th Fusion Energy Conference (Kyoto, Japan, 17-22 October 2016)}
\begin{abstract}
	Quasilinear turbulent transport models are a successful tool for prediction of core tokamak plasma profiles in many regimes. Their success hinges on the reproduction of local nonlinear gyrokinetic fluxes. We focus on significant progress in the quasilinear gyrokinetic transport model QuaLiKiz [C. Bourdelle \textit{et al.} 2016 \textit{Plasma Phys. Control. Fusion} \textbf{58} 014036], which employs an approximated solution of the mode structures to significantly speed up computation time compared to full linear gyrokinetic solvers. Optimization of the dispersion relation solution algorithm within integrated modelling applications leads to flux calculations $\times10^{6-7}$ faster than local nonlinear simulations. This allows tractable simulation of flux-driven dynamic profile evolution including all transport channels: ion and electron heat, main particles, impurities, and momentum. Furthermore, QuaLiKiz now includes the impact of rotation and temperature anisotropy induced poloidal asymmetry on heavy impurity transport, important for W-transport applications. Application within the JETTO integrated modelling code results in 1~s of JET plasma simulation within 10 hours using 10 CPUs. Simultaneous predictions of core density, temperature, and toroidal rotation profiles for both JET hybrid and baseline experiments are presented, covering both ion and electron turbulence scales. The simulations are successfully compared to measured profiles, with agreement mostly in the 5-25\% range according to standard figures of merit. QuaLiKiz is now open source and available at www.qualikiz.com.  
\end{abstract}
\maketitle

\section{Introduction}
\label{sec:section1}
An accurate and predictive model for turbulent transport fluxes driven by microinstabilities is a vital component of predictive tokamak plasma simulation. This enables the interpretation and optimization of present-day experiments, extrapolation to future machine performance, and design of control systems. 

We report on significant progress made in the tractability and validation of the quasilinear gyrokin\textsl{}etic transport model QuaLiKiz. The basis of the QuaLiKiz model is extensively described in Refs.~\cite{bour07,casaphd,bourhdr,bour16}. 

Optimization of the dispersion relation solver has accelerated the calculation time by a factor 50 compared to the QuaLiKiz version reported in Refs~\cite{cott14,bour16}. The dispersion relation for a single wavenumber is now solved within $\sim$1~s. This allows tractable simulation of flux-driven dynamic profile evolution including all transport channels: ion and electron heat, main particles, impurities, and momentum. QuaLiKiz is now of comparable speed as the TGLF gyrofluid quasilinear model~\cite{stae07,kins08}, used for similar multi-channel applications~\cite{stae14}. These numerical improvements are summarized in Section~\ref{sec:section2} and \ref{app:opt}.

Additional physics has been added, widening the applicability of the model. This includes poloidal asymmetries, important for heavy impurity transport, reproducing previous results obtained with GKW~\cite{angi12}. The ETG model has been recalibrated, and includes a rudimentary multi-scale model. The eigenfunction solver has been improved, resulting in further validation of the impact of $E{\times}B$ shear and momentum transport. These additions are summarized in Section~\ref{sec:section3}.

QuaLiKiz has been coupled to the JETTO-SANCO~\cite{cena88,roma14} integrated modelling suite. The code speed-up now allows, for the first time with QuaLiKiz, tractable heat, particle, and momentum transport simulations for main ion, impurity, temperature, and toroidal rotation core profile predictions. Validation was carried out on JET hybrid and baseline pulses. Agreement with measured profiles were mostly in the 5-25\% range according to standard figures of merit. The integrated modelling simulations are described in Section~\ref{sec:section4}.

The motivation for development of quasilinear transport models such as QuaLiKiz is the prohibitive computational cost of direct numerical simulation with massively parallel nonlinear gyrokinetic codes over discharge timescales. Even for gradient-driven, ${\delta}f$ simulations using codes such as \textsc{Gene}~\cite{gene}, GKW~\cite{gkw}, GYRO~\cite{gyro}, and GS2~\cite{gs2}, the computational cost for either ion or electron scale simulations is $\sim10^4$ CPUh for fluxes at a single radius, and is significantly larger for multi-scale simulations~\cite{howa16}. Assuming $O(10^3)$ flux calls per 1s of JET-scale plasma evolution, and 10 radial points, the computational cost for 1s of discharge profile evolution is $\sim10^8$~CPUh. Use of implicit time-step solvers~\cite{barn10b} or steady-state iteration schemes~\cite{cand09} can reduce the computation time, and provides unique opportunities for validation of nonlinear predictions. Nevertheless, routine application of nonlinear turbulence codes for core profile prediction remains elusive, especially when including both ion and electron turbulence scales. 

The quasilinear approximation provides significantly increased tractability. This has proven to be a successful tool for tokamak turbulence model reduction, successfully reproducing nonlinear simulations over a large range of parameters (see Ref.\cite{bour16} and references therein, and Refs~\cite{stae07,stae13,stae16}). The quasilinear approximation has also been successfully applied for stellarator turbulence modelling~\cite{pues16}. The approximation is valid in the plasma confinement zone where the density fluctuations are small - ${\delta}n/n{\sim}O(\%)$. Their success hinges on the reproduction of local nonlinear gyrokinetic fluxes~\cite{casa09}. 

The specific assumptions made in the QuaLiKiz linear gyrokinetic solver lead to a $\sim$6 order of magnitude computational speed-up compared to local ${\delta}f$ nonlinear gyrokinetics. This brings the computational cost of 1s of JET-scale plasma evolution down to $\sim100$~CPUh, which is routinely feasible. 
These assumptions are as follows: 
\begin{itemize}
	\item Shifted circle ($\hat{s}-\alpha$) geometry with a small inverse aspect ratio expansion. This allows the bounce averaging of the trapped species to be expressed as elliptic integrals, speeding up the dispersion relation calculation time. Our neglect of shaping parameters such as triangularity and elongation is expected to lead to systematic overpredictions of the transport level. However, in the plasma core for standard tokamak shapes, only a relatively minor flux overprediction $(<\sim30\%)$ and critical stability threshold underprediction $(<\sim10\%)$ are expected~\cite{kins07,jenk01}. While the strong dependence of H-mode pedestal, and hence global confinement, on plasma shaping is well established~\cite{leon07}, pedestal confinement is out of the scope of QuaLiKiz model. In the integrated modelling discussed here, we apply the measured pedestal height as boundary conditions for core modelling with QuaLiKiz. For more extreme plasma shapes, for example the low aspect ratio typical of spherical tokamaks, or high ($\kappa>\sim2$) elongation, we do not expect QuaLiKiz to be valid. Generalization to a more complete geometry description is the subject of future work.
	\item Electric potential fluctuations only. To save computation time, we only solve the Poisson field equation (quasineutrality) and not Ampere's Law. This is equivalent to assuming $\beta=0$. The relevance of QuaLiKiz for spherical tokamak configurations is thus further reduced, since microtearing modes can dominate the electron heat transport in high-$\beta$ regions~\cite{roac05,terr15}. Finite-$\beta$ stabilization of ITG is also not included in QuaLiKiz, which can be important for high performance scenarios~\cite{citr15a,garc15,doer16}. However, since the nonlinear stabilization (flux reduction) of ITG due to finite-$\beta$ is greater than would be expected by the linear $\beta$-stabilization~\cite{citr13,pues13b}, future work will explore modifying the QuaLikiz saturation rule to capture this nonlinear effect.
	\item Shifted Gaussian (ballooned) eigenfunction ansatz. A significant gain in computational speed ($\sim$2 orders of magnitude) is achieved by a non-self-consistent solution of the eigenvalue (real frequency and growth rate) and eigenfunction (radial and poloidal structure of the electrostatic potential) in the dispersion relation. The eigenfunction is first calculated through expanding the dispersion relation under a strongly driven assumption. See \ref{app:fluid} for further details. This eigenfunction is then inserted into the kinetic dispersion relation which is solved using a weak formulation of Poisson's equation~\cite{bourhdr}. The eigenfunction ansatz is restricted to strongly ballooned Gaussian structures. This limits the dispersion relation solution accuracy in cases where broad poloidal eigenfunctions are observed in self-consistent calculations. This limits the accuracy of trapped electron mode calculations. Furthermore, slab-like ITG is not captured, as observed at low magnetic shear or significant Shafranov shift. For examples, see Ref.~\cite{citr12} and section~\ref{sec:oliver} of this paper. However, in spite of its relative simplicity, the symmetry breaking induced by the shifted Gaussian mode structure in the presence of rotation has proven successful for QuaLiKiz calculations of momentum transport and $E{\times}B$ suppression of transport~\cite{cott14}.
\end{itemize}

\section{Dispersion relation solver optimization}
\label{sec:section2}
The QuaLiKiz calculation is comprised of three steps. The first step is the eigenvalue solver that calculates the instability frequencies and growth rates. The second step calculates the quasilinear flux integrals. These two steps are typically carried out for a spectrum of unstable modes. The third step is the saturation rule. This builds the final transport fluxes by combining the spectrum of quasilinear flux integrals with a modelled spectral form factor and amplitude level based on nonlinear simulations. 

The bulk of time spent for the QuaLiKiz calculation is in the first step, the eigenvalue solver. This is a root finding algorithm in the complex plane, of the dispersion relation $D(\omega)=0$ found by setting the quasineutrality constraint on the perturbed plasma densities from the linearised and gyroaveraged Vlasov equation. See Refs.~\cite{bour16,bourhdr} for further details of the form of $D(\omega)$ solved in QuaLiKiz. $\omega$ is the eigenvalue $\omega_r+i\gamma$, where $\omega_r$ is the real frequency and $\gamma$ the growth rate. The QuaLiKiz algorithm only searches for solutions in the upper half of the complex plane, i.e. for unstable modes (positive $\gamma$), and not for damped modes (negative $\gamma$). 

The QuaLiKiz eigenvalue solver algorithm has been optimized. This led to a calculation speed-up by an approximate total factor of $50-100$. This was key to transform QuaLiKiz into a pragmatic tool for integrated modelling applications. Per wavenumber, the QuaLiKiz calculation time is now $\sim0.5-1~s$ when not including rotation, and $\sim2-4~s$ when including rotation (longer, due to a loss of symmetry in the integrations). We briefly summarize the various steps taken for this optimization. Full details are found in \ref{app:opt}.

\begin{itemize}
	\item The root finding algorithm in the QuaLiKiz eigenvalue solver follows the Davies method~\cite{davi86}. The contour paths have been extensively optimized, minimizing the number of individual $D(\omega)$ calculations. This included modifying the contour shapes, contour overlap, and contour ranges. A significant speed-up of factor $\sim$8 is achieved compared to the previous QuaLiKiz version. A similar level of robustness in the root finding is achieved as in the previous version, since we maintain full coverage of the physically relevant regions in the complex plane.
	\item Improved numerical techniques for integral functions within $D(\omega)$. For the plasma dispersion function, the standard ``WofZ''  method~\cite{gaut69} was replaced by the faster Weideman algorithm~\cite{weid94}. Furthermore, the Carlson method~\cite{carl95} for calculating elliptic integrals was replaced by the T. Fukushima method~\cite{fuku15}. Together, a further factor $\sim3$ speed-up was achieved compared to the previous QuaLiKiz version.
	\item Within integrated modelling simulations, at a given timestep, the eigenvalue solution from the previous timestep is used to accelerate the calculation. The previous solution is used as a first guess for a Newton solver to converge to the new eigenvalues. The validity of this approach is through the relatively small modification in background plasma profiles between the $\Delta{t}\sim1ms$ timesteps typical of integrated modelling simulations. At every $\approx10~ms$, the full contour solution is again sought out, since new eigenvalues may appear. This procedure accelerates the integrated modelling simulation by a factor $\sim5$.
\end{itemize}

\section{Extensions to the QuaLiKiz physics model}
\label{sec:section3}
The physics content in QuaLiKiz has been extended and further validated. This includes an improved eigenfunction solver in the presence of rotation, further validation of momentum transport, allowing an arbitrary number of active or trace ion species, poloidal asymmetry (important for heavy impurity transport), and a new saturation rule for ETG turbulence. These points will be discussed in this section, together with a discussion of caveats and limitations of the $\alpha$-stabilization and $E{\times}B$ shear turbulence suppression model in QuaLiKiz.

\subsection{New eigenfunction solver}
A key feature of QuaLiKiz is the use of an eigenfunction calculated from an expansion of the linear gyrokinetic equation arising from a strongly driven (mode frequency ordered larger than drift and transit frequencies) assumption. This allows for significantly faster eigenvalue calculations. This approximated eigenfunction has been compared to self-consistent gyrokinetic eigenfunctions calculated by \textsc{Gene}, GKW, GYRO, and agrees in regimes where the eigenfunction is ballooned~\cite{citr12,cott14}. 

Including the eigenfunction shift due to rotation effects breaks the eigenfunction symmetry leading to angular momentum flux. The success of applying approximated shifted eigenfunctions for momentum transport predictions is not trivial. This approach in QuaLiKiz has been validated through comparisons against angular momentum flux predictions with non-approximated eigenfunctions~\cite{cott14}. 

The fluid eigenfunction solution in the present version of QuaLiKiz has been rederived and improved compared with Ref.~\cite{cott14}. Particular care has been taken in setting up a consistent and less restrictive ordering of the various terms. The model is also refined by consistently carrying out pitch angle integrations. An extensive overview of the revised derivation is presented in \ref{app:fluid}. 

\subsection{Verification of momentum transport model} 
\begin{figure}[htbp]
	\centering
	\includegraphics[scale=0.8]{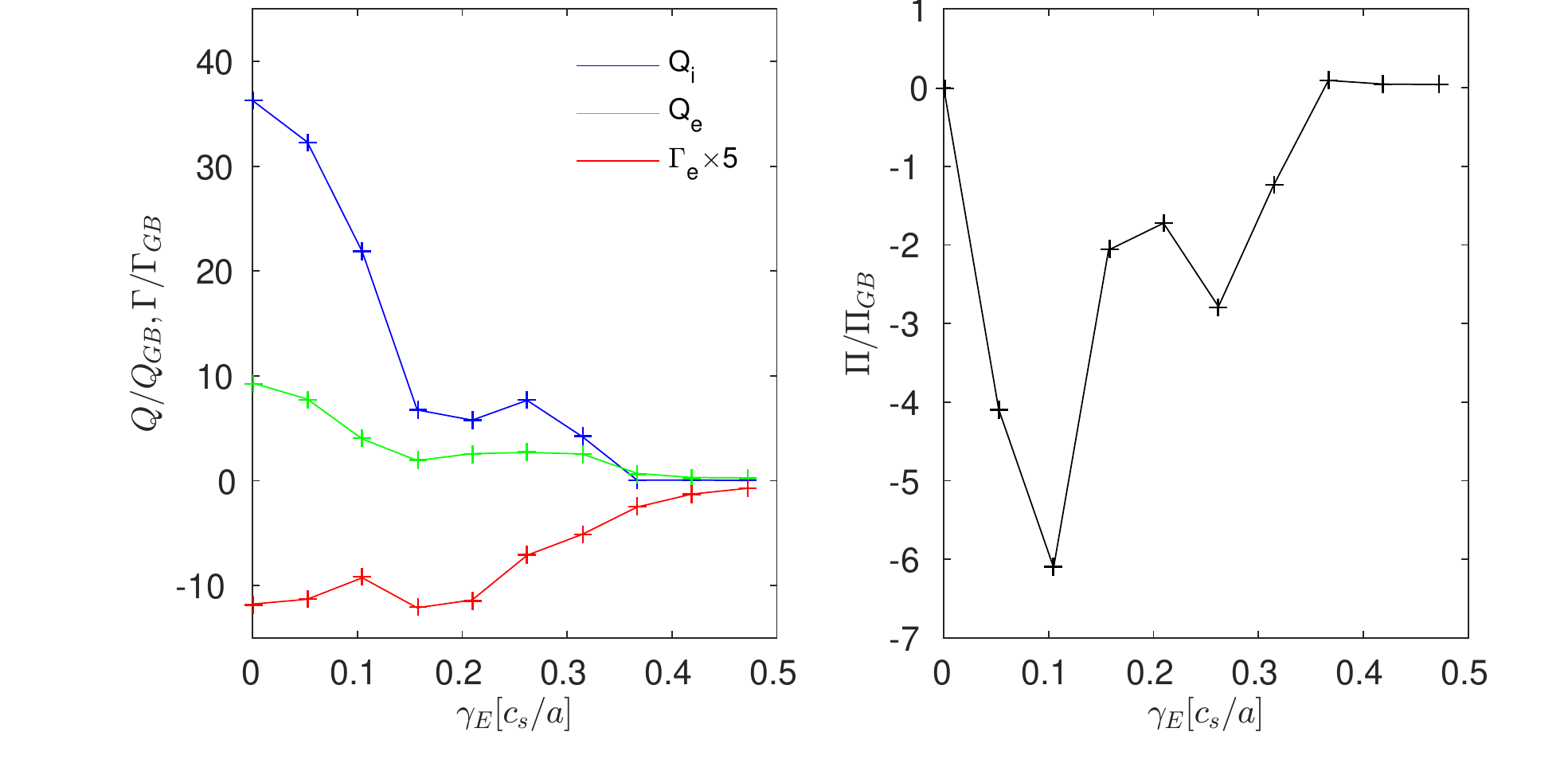}
	\caption{\footnotesize Transport fluxes (left panel) and momentum transport (right panel) as a function of perpendicular velocity shear $\gamma_E\equiv-\frac{dv_\perp}{dr}$, with $M=M'=0$. $\gamma_E$ is normalized to $c_s/a$, and the transport fluxes are normalized to GyroBohm flux units. $Q_{sGB}\equiv\frac{T_s^{2.5}m_i^{0.5}n_s}{r^2B^2e^2}$, $\Gamma_{GB}\equiv\frac{T_s^{1.5}m_i^{0.5}n_s}{r^2B^2e^2}$, and $\Pi_{GB}\equiv\frac{T_s^{2}n_sm_iR}{r^2B^2e^2}$}
	\label{fig:figgammaescan}
\end{figure}

Verification of the QuaLiKiz momentum transport model using the new eigenfunction solver is reported here.  First, we show that we recover previous predictions of the perpendicular $E\times{B}$ shear impact on turbulence suppression and momentum transport. This is seen in figure~\ref{fig:figgammaescan}, and carried out for GA-Standard-Case (GASTD) parameters, with 8 toroidal modes at ITG length-scales ($k_\theta\rho_s=\left[0.1,0.175,0.25,0.325,0.4,0.5,0.7,1.0\right]$), where $\rho_s\equiv\sqrt{T_em_i}/Z_iB_0$. Similarly to both Ref.~\cite{cott14} (Fig. 9 therein) and Ref~\cite{stae13} (Fig 4), the impact of $\gamma_E$ on the fluxes leads to full suppression at approximately $\times2$ the maximum linear growth rate, and a similar degree of $\gamma_E$ driven momentum flux. 

\begin{figure}[htbp]
	\centering
	\includegraphics[scale=0.8]{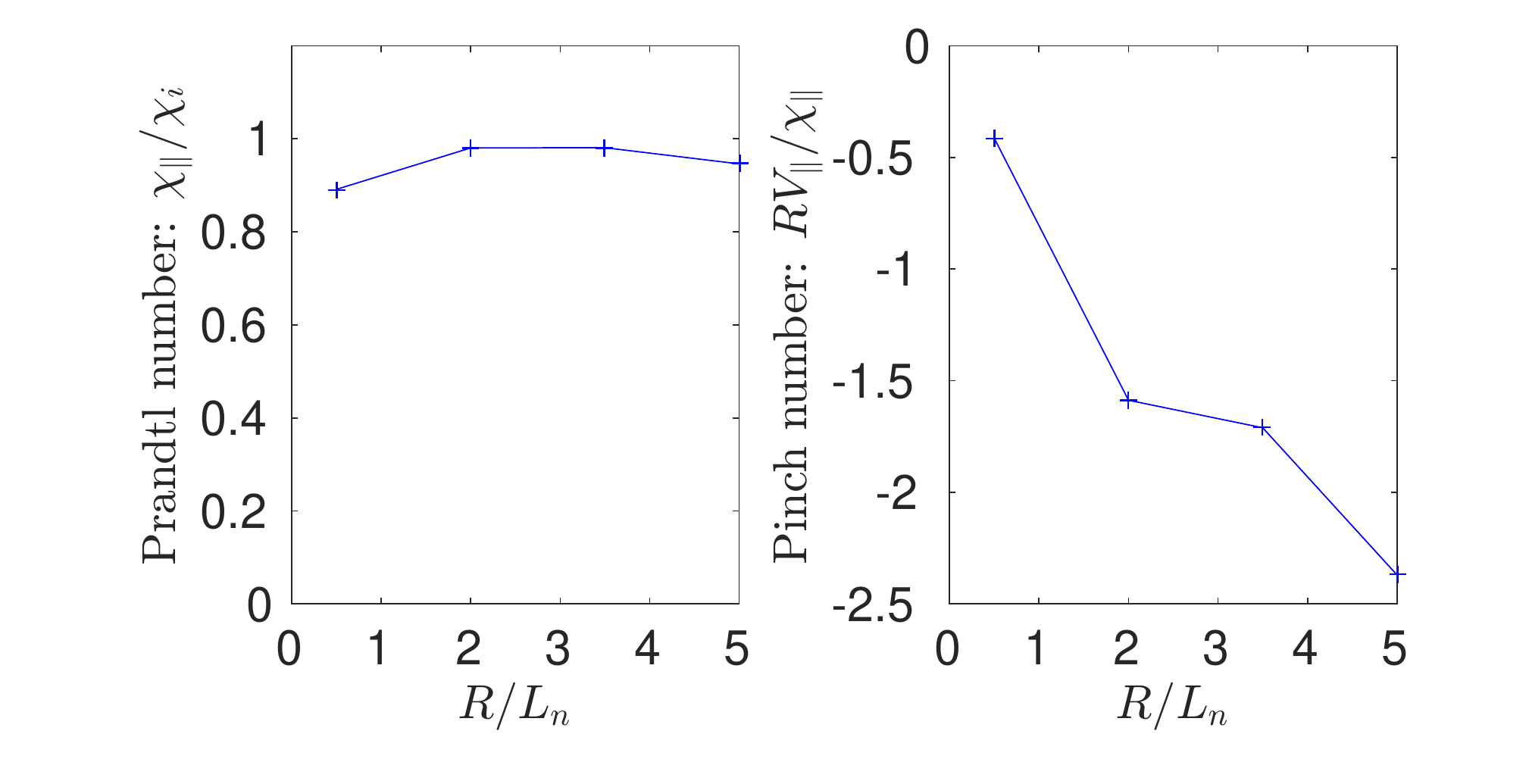}
	\caption{\footnotesize Scan of impact of trapped electron drive on QuaLiKiz momentum transport predictions via a $R/L_n$ scan based on GASTD case parameters. The Prandtl number is shown in the left panel, and the Pinch Number in the right panel.}
	\label{fig:figprandtlscan}
\end{figure}

QuaLiKiz also predicts increasing momentum pinch with trapped electron drive, as predicted by analytic theory and gyrokinetic simulations~\cite{peet11}. In figure~\ref{fig:figprandtlscan}, we show the Prandtl Number $\equiv\chi_{\parallel}/\chi_i$ (left panel) and momentum Pinch Number ${\equiv}RV_\parallel/\chi_\parallel$ (right panel) for a GASTD case $R/L_n$ scan. $\chi_i$ is the ion heat diffusivity. $\chi_\parallel$ is the momentum diffusivity, $V_\parallel$ is the momentum pinch, as defined by the momentum flux decomposition:
\begin{dmath}
\Pi_\parallel=R\sum_sm_sn_s\left(-\chi_{s\parallel}\frac{du_{s\parallel}}{dr}+V_{s\parallel}u_{s\parallel}\right)+\Pi_{RS}
\end{dmath}
where $\Pi_{RS}$ is the residual stress. Note that the GASTD case has a single ion species, so the summation is not relevant here. The pinch is calculated from carrying out the $R/L_n$ scan with $M{\equiv}u_\parallel/v_{thi}=0.2$ and $M'=0$. The momentum diffusivity is calculated from a $R/L_n$ scan with $M=0$ and $M'=1$, where $M'\equiv-R\frac{dM}{dr}$. As expected, while the Prandtl Number is nearly constant, the Pinch Number increases (in absolute terms) with increasing $R/L_n$, i.e. increasing trapped electron drive. This scan is identical to that shown in Ref.~\cite{cott14} (see figure 10 therein). However, with the new eigenfunction model, the Prandtl number is now closer to unity, and the Pinch Number scan is upshifted by approximately 1.5 units, both in disagreement with full gyrokinetic modelling (see figures 5 and 7 in Ref.~\cite{peet11}). Nevertheless, the new eigenfunction model assumptions and ordering are more consistent compared to that presented in Ref.~\cite{cott14}. Therefore, we maintain the new model in spite of the reduced accuracy in comparison to full gyrokinetic modelling in this particular scan.

\begin{figure}[htbp]
	\centering
	\includegraphics[scale=0.8]{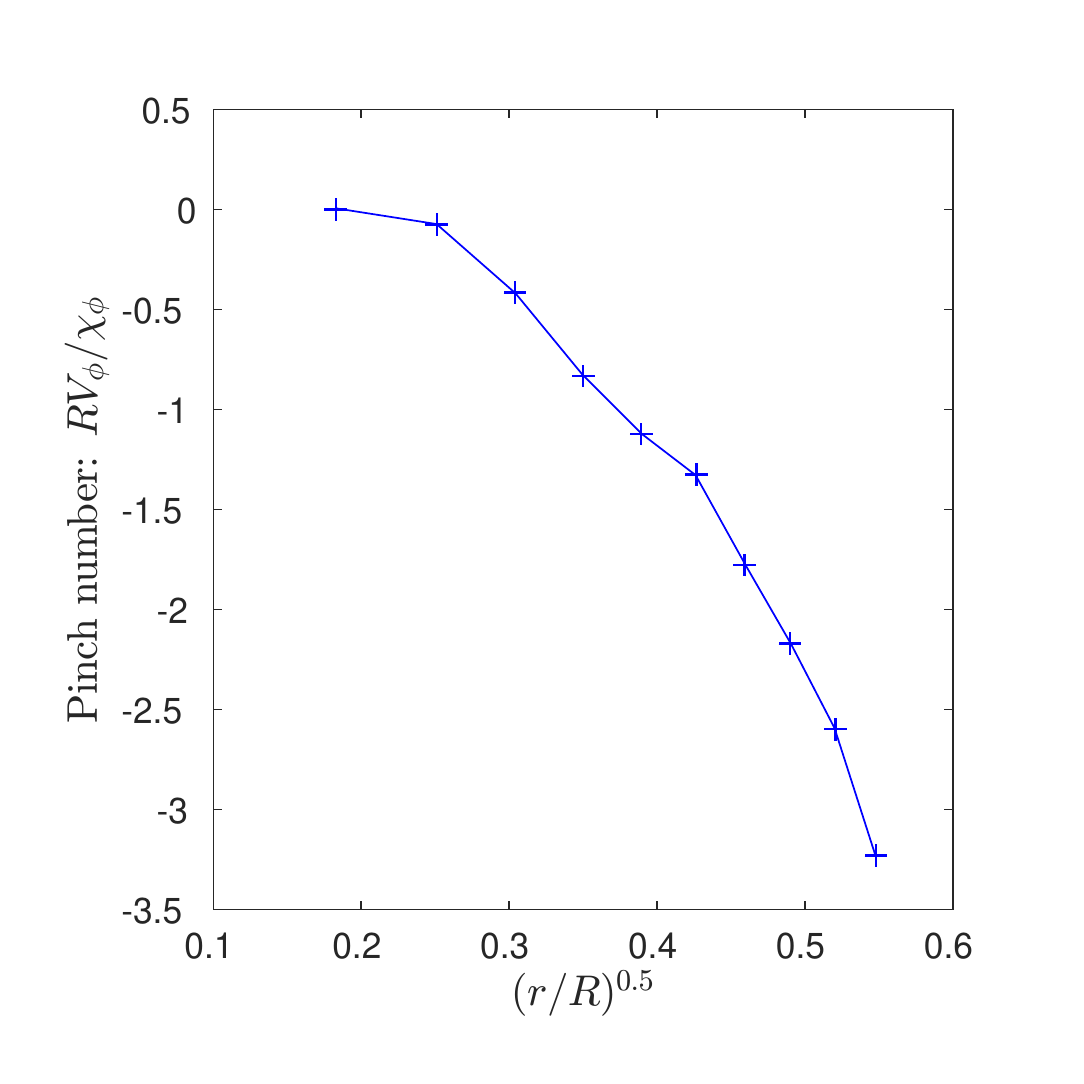}
	\caption{\footnotesize Scan of impact of trapped electron drive on QuaLiKiz momentum pinch predictions via a $r/R$ scan based on GASTD case parameters}
	\label{fig:figepsscan}
\end{figure}

Finally, beyond the scope of the previous studies~\cite{cott14}, we also explored the sensitivity of the Pinch Number to local inverse aspect ratio $\epsilon{\equiv}r/R$, in an $\epsilon$ scan based on GASTD case parameters. This is effectively a scan of the trapped electron drive, due to the proportionality of the trapped electron fraction to $\epsilon^{0.5}$. The results are shown in figure~\ref{fig:figepsscan}. As expected, a clear increase in momentum pinch is observed with increasing $(r/R)^{0.5}$. This further corroborates the QuaLiKiz rotation model. 

\subsection{Poloidal asymmetries for heavy impurities}
The constraint of parallel force balance in the presence of centrifugal forces, and/or temperature anisotropy (e.g. due to NBI and ICRH heating), leads to asymmetry of particle density $n_j$ on a flux surface~\cite{hint85,cass10,rein12}. Following the Bi-Maxwellian temperature anisotropy model of Ref.~\cite{bila14}, we obtain, at a given radius, for species $j$, the poloidal density profile $n_j(\theta)$:

\begin{dmath}
	\label{eq:force}
	n_j(\theta)=n_{j,LFS}\frac{T_{\perp{j}}(\theta)}{T_{\perp{j,LFS}}}exp\left[-\frac{Z_j\Phi(\theta)-\frac{1}{2}m_j\Omega^2\left(R(\theta)^2-R_{LFS}^2\right)}{T_j}\right]
\end{dmath}  
with:
\begin{dmath}
	\label{eq:Tperp}
	\frac{T_{\perp{j}}(\theta)}{T_{\perp{j},LFS}}=\left[\frac{T_{\perp{j},LFS}}{T_{\parallel{j},LFS}}+\left(1-\frac{T_{\perp{j},LFS}}{T_{\parallel{j},LFS}}\right)\frac{B_{LFS}}{B(\theta)}\right]^{-1}
\end{dmath}  
where $LFS$ denotes the low field side values, $\Phi(\theta)$ is the equilibrium electrostatic potential, $\Omega$ the angular toroidal velocity, $T_{\perp}$ the perpendicular temperature, and $T_{\parallel}$ the parallel temperature. $\Phi(\theta)$ is solved numerically at each poloidal location by invoking the quasineutrality constraint. This has now been implemented in QuaLiKiz and successfully verified against the similar quasineutrality solver in GKW. The inputs into the system are the toroidal velocity, temperature anisotropy at any single poloidal location (here defined as $LFS$ for convenience, but this can be generalized), and density ratios at any single poloidal location (similarly defined here as the $LFS$ for convenience). 

While the density asymmetry is small for light ions, it can be non-negligible for heavy impurities. The high mass leads to significant centrifugal forces, while the high charge leads to a significant response to the equilibrium potential. The density asymmetry can significantly modify the resultant diffusion and pinch for both turbulent and neoclassical transport, and is a critical effect to include for heavy ion impurity transport predictions such as for W~\cite{angi14,angi15,cass15}. 

For turbulent transport in QuaLiKiz, we closely follow the model of Ref.~\cite{angi12}. The impact of the density asymmetry is not included self-consistently in the dispersion relation, which maintains the particle densities as flux functions. However, when evaluating the quasilinear particle flux integrations, we include the density asymmetry. The asymmetries lead to a poloidally dependent density gradient, leading to new terms in the quasilinear integral which effectively emerge as new pinch terms, which can become significant for heavy impurities. These terms are then evaluated by flux surface averaging, and are additionally weighted by the eigenfunction shape. Full details are found in \ref{app:asym}. The flux function inputs into the QuaLiKiz dispersion relation are taken from the $LFS$ values. At low impurity densities (trace values) this choice is of negligible importance. However, for non-trace heavy impurities this choice can have an impact on the dispersion relation predictions. Due to the $\frac{n_sZ_s^2}{T_s}$ prefactors in the dispersion relation, even experimentally relevant low densities of high-Z impurities can have a non-negligible impact on the instability calculation. Future work will focus on how best to approximate the 0D flux function density and density gradient inputs based on the 1D poloidal profile, for such non-trace cases.

\begin{figure}[htbp]
	\centering
	\includegraphics[scale=0.8]{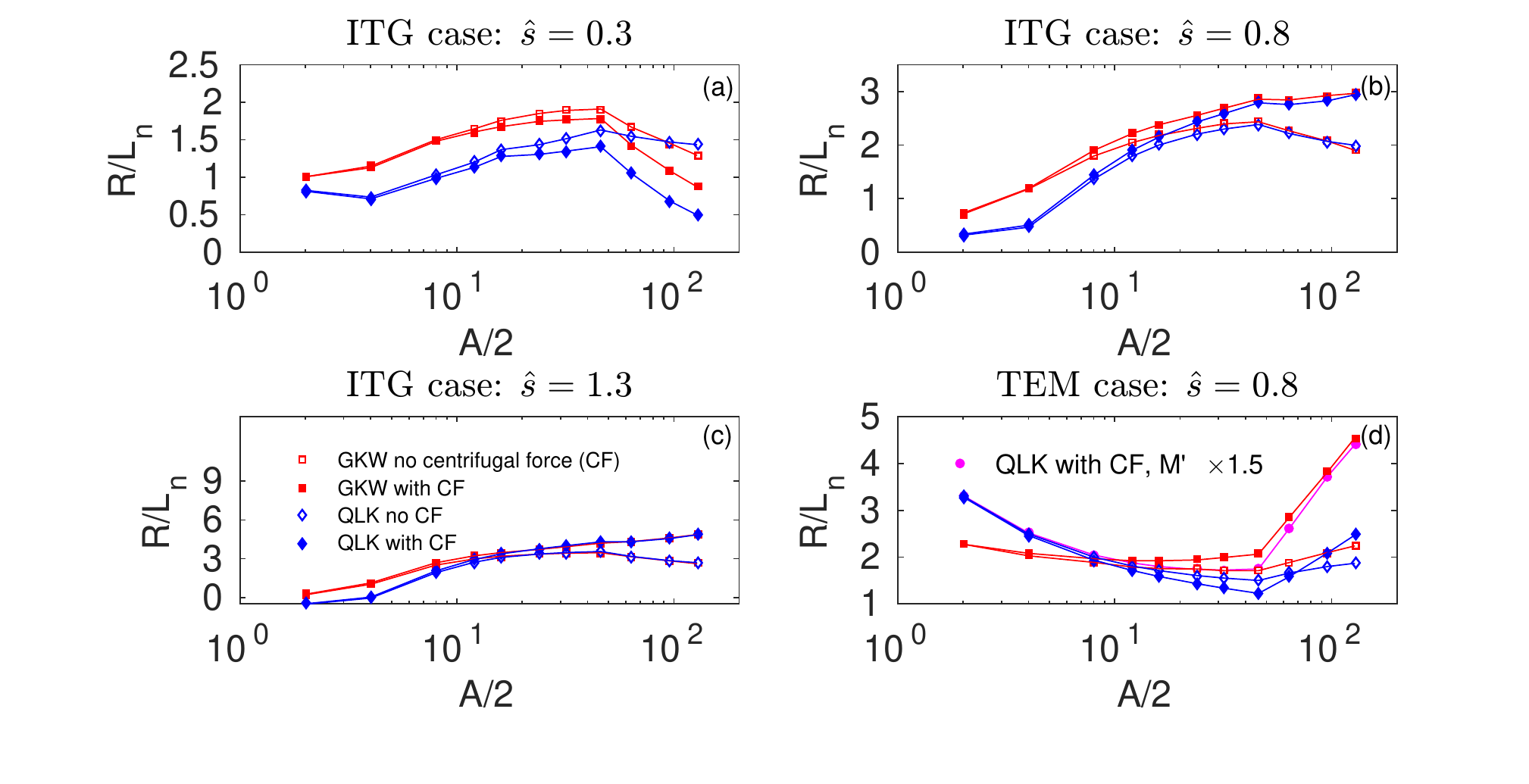}
	\caption{\footnotesize Validation between QuaLiKiz and GKW predictions of impurity $R/L_n$ (assuming no sources) for an ITG dataset at various magnetic shear (panels a-c), and for a TEM dataset (panel d). GKW data is depicted by the square symbols, and QuaLiKiz data by the diamond symbols. In panel d, the curve with circular symbols signifies a QuaLiKiz run with increased M'. The curves including poloidal asymmetry due to centrifugal effects are shown with solid symbols, and when neglecting poloidal asymmetries, with open symbols}
	\label{fig:angioni}
\end{figure}

The QuaLiKiz poloidal asymmetry model for heavy impurity transport was validated against quasilinear GKW predictions, using parameter sets defined in Ref.~\cite{angi12}. A comparison was made of the predicted impurity normalized logarithmic density gradient $R/L_{n}$. Assuming no core sources, this is provided by the zero-flux constraint: $R/L_n=-\frac{RV}{D}$, where $V$ is the convective velocity (pinch) and $D$ the diffusivity.  Collisionless runs were carried out, scanning over the impurity mass A (in amu). The impurity charge throughout the scan was set at $Z=A/2$ for $A<92$, and $Z=46$ for $A>92$. Two basic datasets were employed: an ITG dataset with $R/L_{Ti}=9, R/L_{Te}=6, R/L_{n}=2, q=1.4, \hat{s}=0.8, T_e/T_i=1$, and a TEM dataset with identical parameters apart from $R/L_{Te}=9, R/L_{Ti}=3, T_e/T_i=1$. $\hat{s}$ is magnetic shear, and $q$ the safety factor. For both cases, we simulate a single wavenumber $k_\theta\rho_s=0.3$. The poloidal asymmetries are from centrifugal effects, with $M=0.2, M'=1$ with respect to the main ion. For the ITG case, we also carried out runs at varying $\hat{s}=0.3,1.3$. The comparison is shown in figure~\ref{fig:angioni}. In panels a-c, the ITG comparison is shown. QuaLiKiz captures the GKW predicted $R/L_n$ with an rms error of $\left(\Delta\frac{R}{L_n}\right)_{rms}\approx0.5$ throughout the dataset. The TEM data has a larger rms error of $\left(\Delta\frac{R}{L_n}\right)_{rms}\approx0.8$, with a significant deviation between the QuaLiKiz and GKW predictions at high $A$ when including poloidal asymmetry effects. This is likely due to the known deficiencies of the QuaLiKiz eigenfunction ansatz in fully reproducing gyrokinetic TEM eigenfunctions~\cite{cott14}. Nevertheless, the GKW behaviour can be well captured by QuaLiKiz following a moderate increase of $M'$ from $1$ to $1.5$, as shown in figure~\ref{fig:angioni}d. This illustrates the sensitivity of the effective 2D rotodiffusion pinch.

\subsection{Quasilinear ETG saturation rule}
\begin{figure}[htbp]
	\centering
	\includegraphics[scale=0.8]{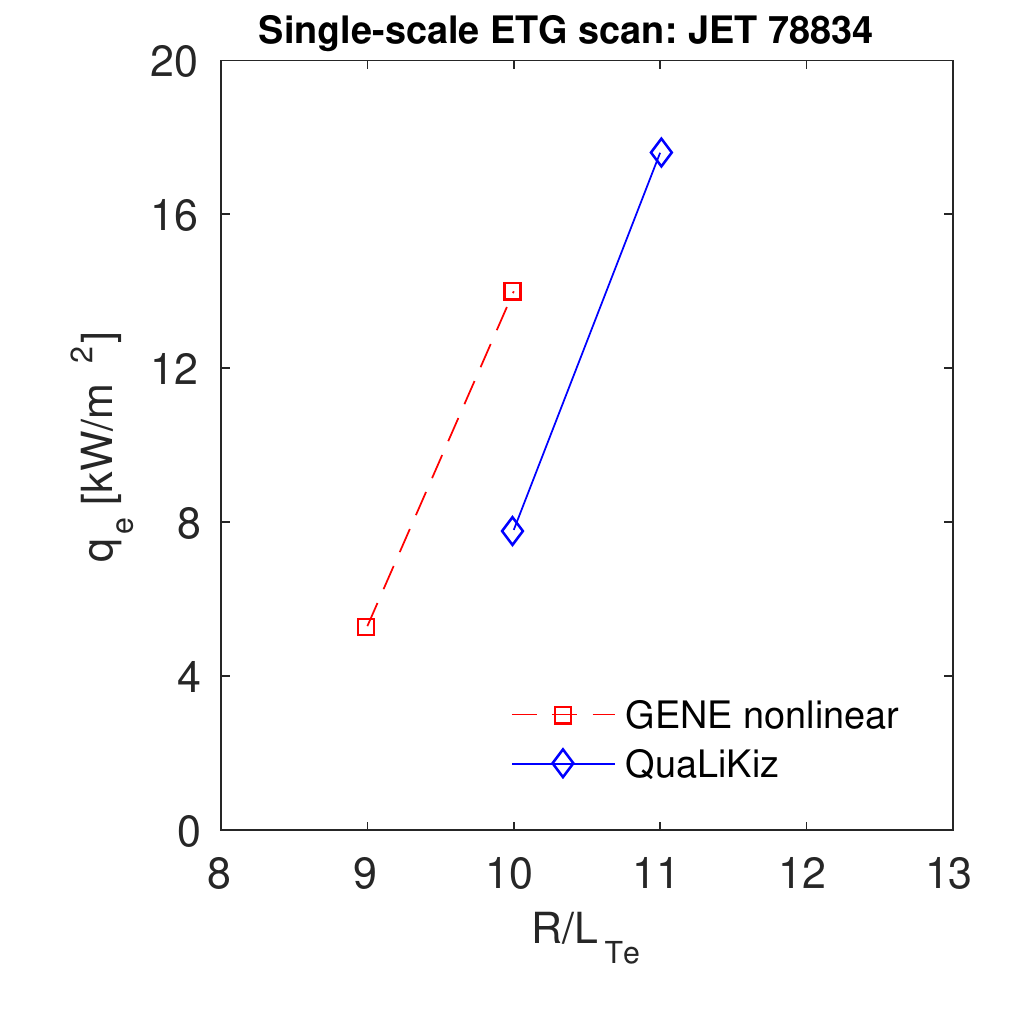}
	\caption{\footnotesize Comparison of QuaLiKiz and nonlinear \textsc{Gene} single-scale ETG simulations, with the new quasilinear ETG QuaLiKiz saturation rule. Red dashed curve is \textsc{Gene} nonlinear simulations, and the solid blue curve is QuaLiKiz. The simulations were carried out at mid-radius, with profiles averaging over 0.5~s of the flattop phase of the discharge}
	\label{fig:ETG}
\end{figure}
The QuaLiKiz ETG nonlinear saturation rule has been recalibrated. Complete isotropisation is assumed, with $k_\perp^2=2k_\theta^2$. The constant prefactor which sets the QuaLiKiz ETG-scale saturation level was tuned by comparison to single-scale ETG nonlinear \textsc{Gene} simulations of JET discharge L-mode discharge 78834~\cite{bona15}. Separate ITG/TEM+ETG single scale nonlinear simulations of this discharge led to heat flux matching in both the ion and electron channels when compared to experimental power balance. Approximately 50\% of the electron heat flux was from ETG turbulence. In the \textsc{Gene} simulations, ETG saturation was achieved by increasing the prescribed level of perpendicular $E\times{B}$ shear until heat flux convergence was reached, due to ETG streamer shearing leading to a reduction of box-scale effects. This technique was assumed to be a proxy for ion-scale eddies shearing the ETG streamers. Since the \textsc{Gene} and QuaLiKiz ETG thresholds differ in this case by approximately 10\%, the calibration was carried out on the stiffness level (gradient of flux with respect to driving gradient). This is seen in figure~\ref{fig:ETG}, where the \textsc{Gene} and QuaLiKiz predictions are parallel. The dimensionless parameters for the simulation input is summarized in table~\ref{tab:ETG}. The dimensional reference parameters for defining the SI fluxes and collisionality were $B_{\mathrm{ref}}=3.5~\mathrm{T}$, $T_{\mathrm{ref}}=1.45~\mathrm{keV}$, $n_{\mathrm{ref}}=1.95\cdot10^{19}~\mathrm{m^{-3}}$, and $L_{\mathrm{ref}}=2.94~\mathrm{m}$. 

\begin{table*}[tp]
	\small
	\centering
	\caption{\footnotesize JET 78834 dimensionless parameters as input into the single-scale ETG nonlinear-\textsc{Gene} and QuaLiKiz $R/L_{Te}$ scan used to recalibrate the QuaLiKiz ETG saturation rule. A carbon impurity was included in the calculation}
	%\vspace{0.15cm}
	\tabcolsep=0.11cm
	\scalebox{0.9}{\begin{tabular}{c|c|c|c|c|c|c}
			\label{tab:ETG}
			$R/L_{Ti}$ & $R/L_{n}$ & $T_i/T_e$ & $r/R$ & $\hat{s}$ & $q$ &  $Z_{\mathrm{eff}}$ \\
			\hline
			5.2 & 2.1 & 0.83 & 0.19 & 1.1 & 2.1 & 1.8 \\
	\end{tabular}}
\end{table*}

The QuaLiKiz ETG saturation rule contains a rudimentary multi-scale effect. The emergence of significant ETG fluxes depends on the relative weight of ion-scale and electron-scale instability. For strongly driven ion-scale modes (significantly above threshold), ETG turbulence is quenched according to observations from multi-scale simulations~\cite{maey15,howa16}. A ``rule of thumb'' based on a dataset of multi-scale simulations, suggests a threshold for non-quenched ETG flux at $\lambda{\equiv}max(\gamma_{ETG}/\gamma_{ITG-TEM})>\sqrt{m_i/m_e}$~ \cite{howa16pop}. $\lambda$ is the growth rate ratio determined by the maxima of the ion-scale and electron-scale instability spectra. Such a rule is set in QuaLiKiz, where the single-scale ETG saturation factor is multiplied by the sigmoid function:
\begin{dmath}
\frac{1}{1+e^{-(\lambda-\sqrt{m_i/m_e})/5}}
\end{dmath}
The sigmoid is employed to ensure smoothness around the transition, to avoid numerical instabilities in integrated modelling frameworks.

This QuaLiKiz ETG model is not valid in regimes with fully absent or suppressed ion-scale turbulence, where significant ETG zonal flows may arise, leading to an extensive ETG threshold shift and strong collisionality dependence~\cite{coly16}. 

Future work will aim towards a more complete QuaLiKiz multi-scale model for ETG saturation, similar to Ref.~\cite{stae16}.

\subsection{Caveats regarding $\alpha$ and $E{\times}B$ stabilization}
\subsubsection{$\alpha$-stabilization}
\label{sec:oliver}
\begin{figure}[htbp]
	\centering
		\includegraphics[scale=0.8]{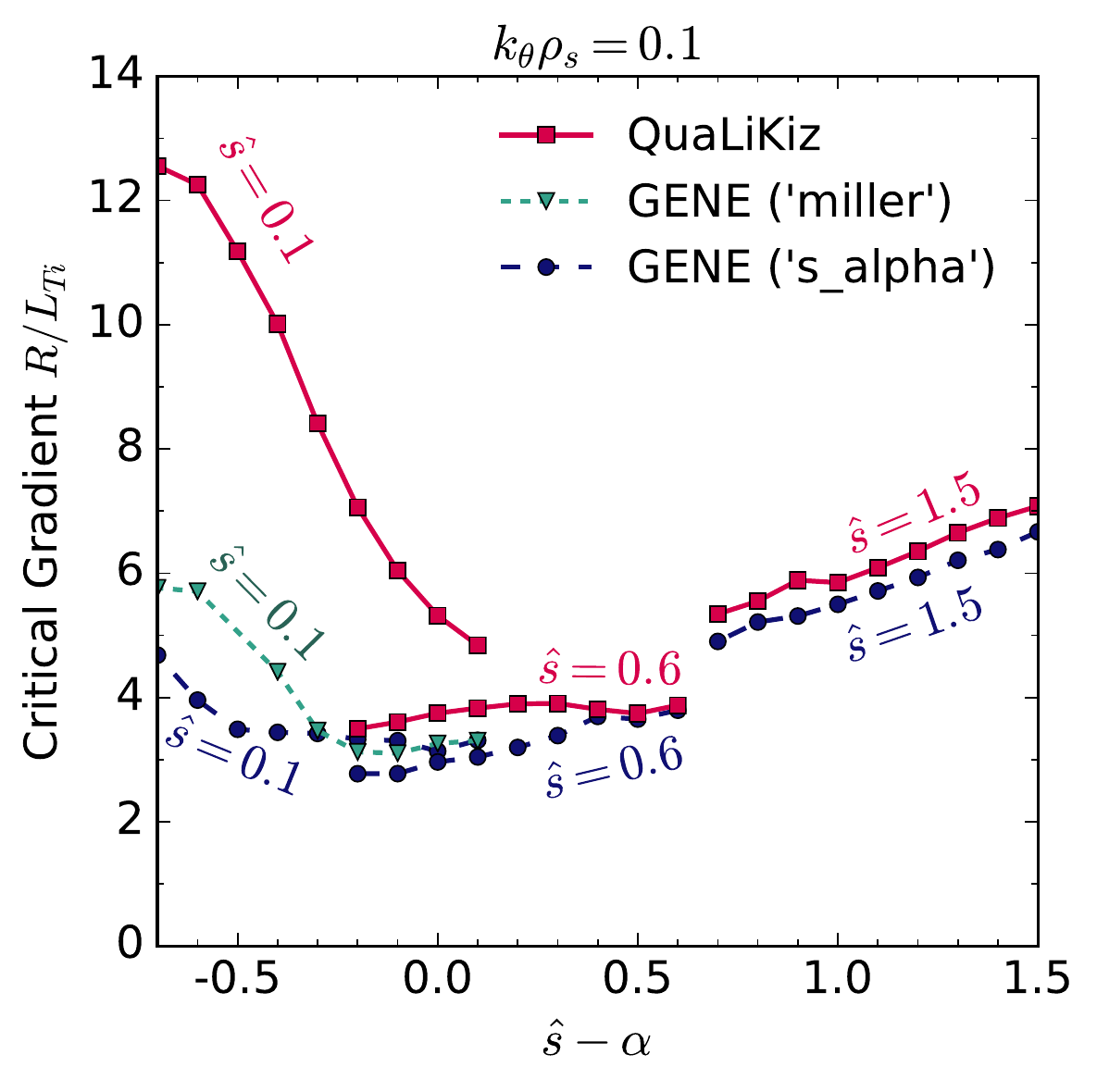}
	\caption{\footnotesize Comparison between ITG threshold predictions of QuaLiKiz (red curves) and linear \textsc{Gene} $\hat{s}-\alpha$ geometry (blue curves) and Miller~\cite{mill98} geometry (green curve) calculations. Various scans around GASTD parameters are carried out. $\alpha$ is scanned in all cases, which differ by $\hat{s}$ value, as labelled in the plot. $k_\theta\rho_s=0.1$ for all cases.}
	\label{fig:oliver1}
\end{figure}

 The validity of the $\alpha$-stabilization model in QuaLiKiz was investigated, by comparison with linear \textsc{Gene} gyrokinetic simulations. This stabilization is due to the modification (and even sign reversal) of the magnetic drift frequency~\cite{bour05}, typically for $\hat{s}-\alpha<0.5$, where $\alpha=q^2\sum_s\beta_s\left(\rlns+\rlts\right)$, and $\beta_s$ is the species dependent ratio of kinetic to magnetic pressure. It was found that QuaLiKiz overestimates the $\alpha$-stabilization for values at approximately $\hat{s}-\alpha<0.2$, particularly at larger wavelengths. A typical example is shown in figure~\ref{fig:oliver1}, based on the GASTD case with $k_\theta\rho_s=0.1$. The critical threshold of the ITG mode is shown, comparing QuaLiKiz predictions to linear-\textsc{Gene} predictions, at various values of magnetic shear ($\hat{s}$) and scanning over $\alpha$. While at higher $\hat{s}-\alpha$, the agreement between the codes is good, at lower $\hat{s}-\alpha$ the QuaLiKiz predicted critical gradient is significantly higher than the \textsc{Gene} prediction. 

\begin{figure}[htbp]
	\centering
		\includegraphics[scale=0.8]{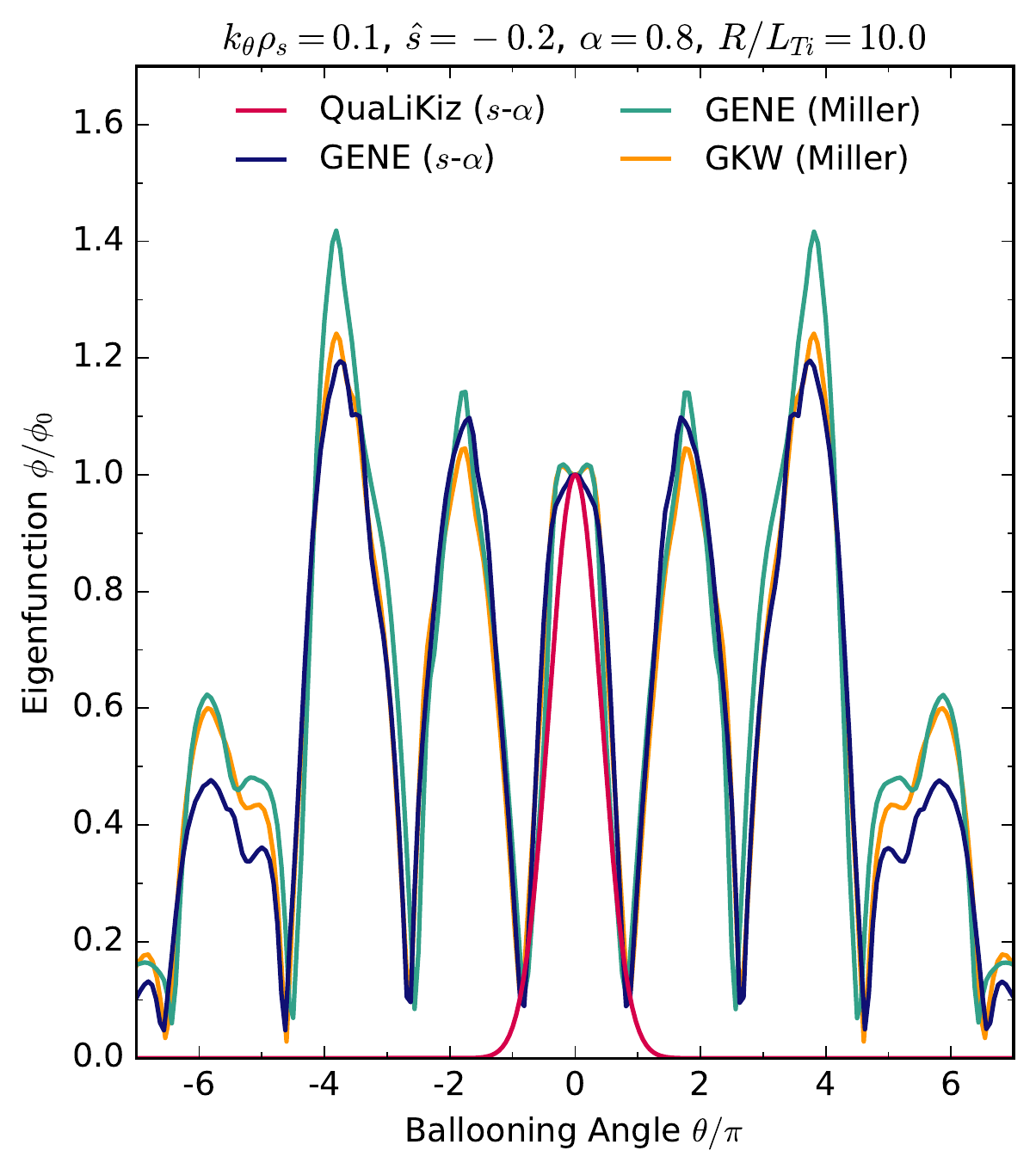}
	\caption{\footnotesize Comparison between QuaLiKiz (red curve), GENE (blue and green curves), and GKW (gold curve) eigenfunction predictions for a GASTD case with $k_\theta\rho_s=0.1$, $\hat{s}=0.2$, $\alpha=0.8$ and $R/L_{Ti}=10$. The QuaLiKiz eigenfunction width is significantly narrower in ballooning space compared to the full gyrokinetic linear predictions, for these parameters.} 
	\label{fig:oliver2}
\end{figure}

This divergence is correlated with a significant disagreement of the eigenfunction predictions between \textsc{Gene} and QuaLiKiz at low-$\alpha$. This is shown in figure~\ref{fig:oliver2}, for a low $\hat{s}-\alpha=-1$ case. The \textsc{Gene} eigenfunction is significantly wider. The wide (in the parallel direction) eigenfunction is a signature of a slab-like mode. Due to the QuaLiKiz ballooned eigenfunction ansatz, these slab modes are not captured in QuaLiKiz. This likely contributes to the disagreement. 

\begin{figure}[htbp]
	\centering
		\includegraphics[scale=0.8]{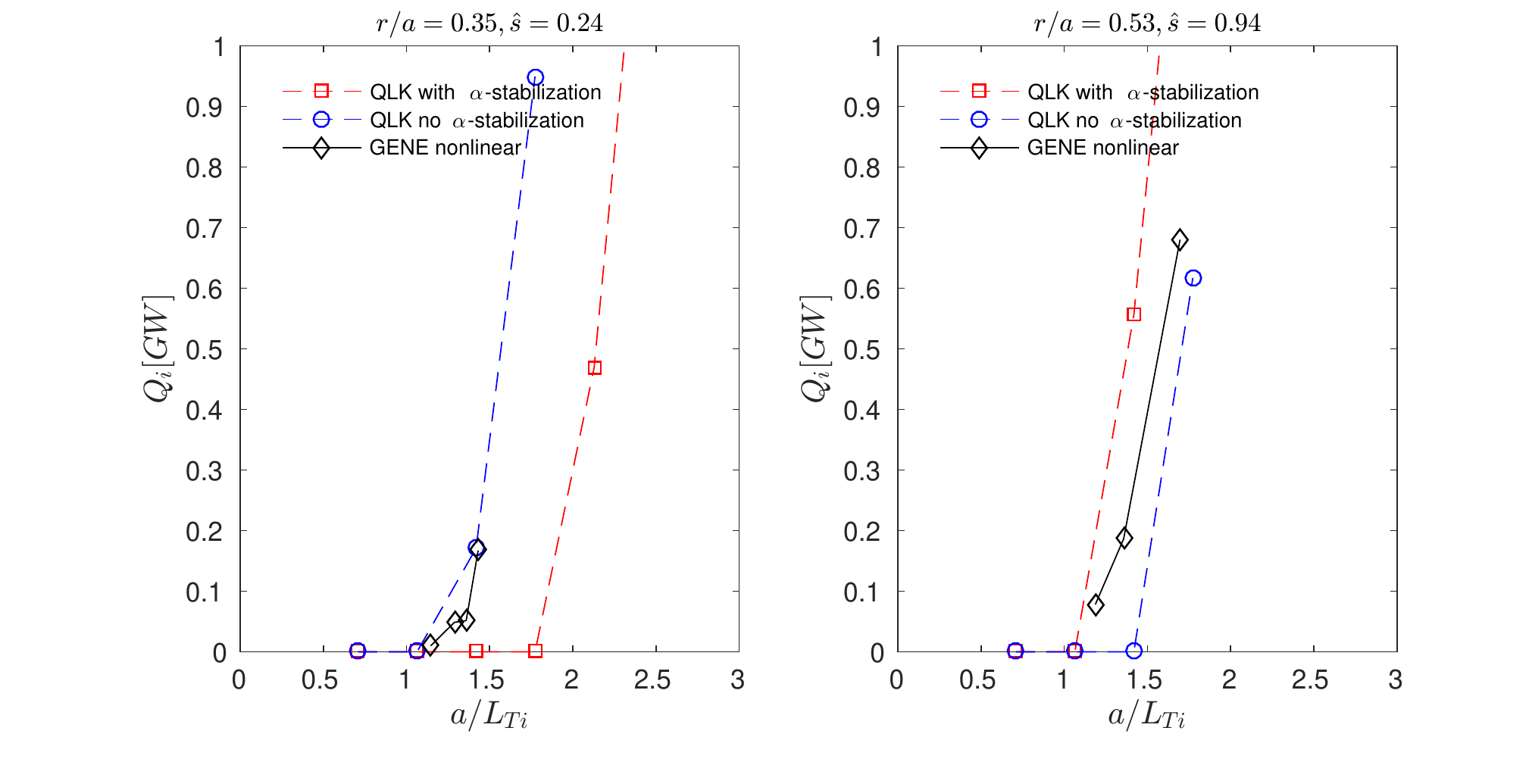}
	\caption{\footnotesize Comparison between nonlinear \textsc{Gene} and QuaLiKiz ion heat flux predictions for DEMO1 parameters, at mid-radius (right panel) and inner-radius (left panel). For both parameter sets, the QuaLiKiz calculations are compared to the \textsc{Gene} results when including (red curves) and neglecting (blue curves) $\alpha$ effects.} 
	\label{fig:demo}
\end{figure}

A further illustration of the validity range of the QuaLiKiz $\alpha$-stabilization model is seen in figure~\ref{fig:demo}. QuaLiKiz and nonlinear-GENE fluxes are compared for DEMO1 flattop parameters~\cite{fabl17,wenn17}. At mid-radius (right panel), the magnetic shear is relatively high, and thus $\hat{s}-\alpha$ values lies within the QuaLiKiz validity range. There, better agreement with the nonlinear \textsc{Gene} turbulence threshold is achieved when including the $\alpha$ model. Note that for these parameters, larger $\alpha$ leads to slight destabilization. However, at low magnetic shear, at more inner radii (left panel), then the low $\hat{s}-\alpha$ means that at finite $\alpha$,  QuaLiKiz significantly overpredicts the ITG instability threshold compared to \textsc{Gene}. Simply setting $\alpha=0$ in QuaLiKiz then leads to significantly improved agreement for this case. 

At low $\hat{s}-\alpha$, the \textsc{Gene} threshold predictions show a relatively weak dependence (see figure~\ref{fig:oliver1}). Therefore, as a workaround to the limitations of the QuaLiKiz model at low $\hat{s}-\alpha$, we modify the inputs. If $\hat{s}-\alpha<0.2$ for $\hat{s}>0.2$, then $\alpha$ is modified such that $\hat{s}-\alpha=0.2$. For $\hat{s}<0.2$, then $\alpha$ is simply modified to 0. This is motivated by the slab-like characteristics of the instability at low $\hat{s}-\alpha$ observed in the \textsc{Gene} calculations, which seem to diminish the impact of the strong toroidal ITG branch stabilization observed in QuaLiKiz. Future work will focus on a more self-consistent solution to this QuaLiKiz limitation. 

\subsubsection{$E\times{B}$-stabilization}
\begin{figure}[htbp]
	\centering
	\includegraphics[scale=0.8]{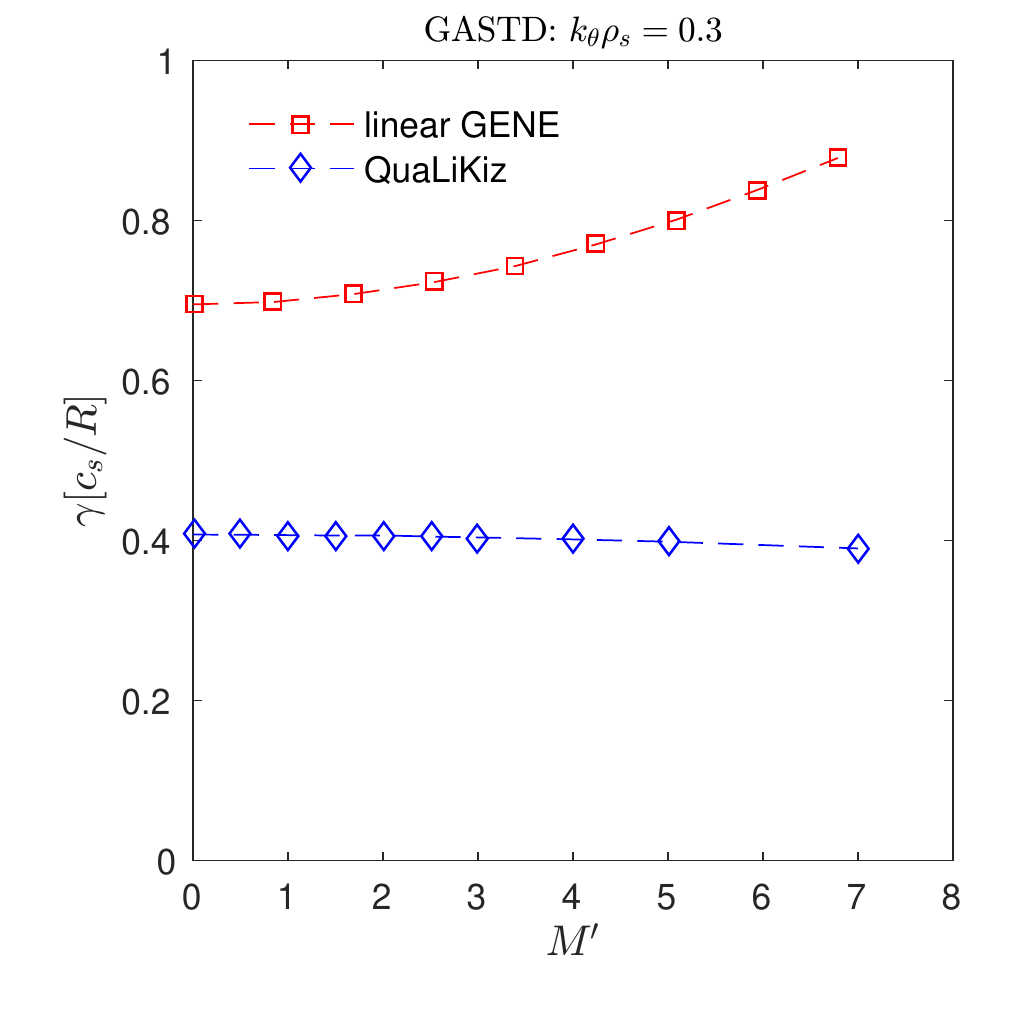}
	\caption{\footnotesize Comparison between linear growth rates calculated by QuaLiKiz (blue curve) and linear-\textsc{Gene} for a M' scan based around the GASTD case with $k_\theta\rho_s=0.3$}
	\label{fig:aupar}
\end{figure}

Parallel velocity gradient (PVG) destabilization can be a significant factor in compensating the $E{\times}B$ flow shear stabilization in tokamaks, particularly at large $q/\epsilon$~\cite{kins05,peet05,cass09,high12,citr14}. However, when using the present QuaLiKiz eigenfunction model, the pure PVG destabilization is not captured, when compared to linear \textsc{Gene} simulations. This is shown in figure~\ref{fig:aupar}, which is a $M'\equiv-\frac{R}{v_{th}}\frac{du_\parallel}{dr}$ scan of linear growth rate predictions at GASTD case parameters with $k_\theta\rho_s=0.3$ and $\alpha=0$. The growth rate spectrum is directly related to the transport fluxes through the quasilinear response and mixing length rules. 
While the \textsc{Gene} results show a $\sim30\%$ increase in linear growth rate when scanning from $M'=0\rightarrow7$, the QuaLiKiz results do not show this increase, and the growth rates even slightly decrease with higher $M'$. This is due to the stabilizing impact of the eigenfunction shift having a stronger impact than the additional driving $M'$ term in the diamagnetic frequency. Note that experimental $M'$ typically does not exceed $M'\sim4$.

This disagreement may have implications for $E\times{B}$ turbulence suppression predictions at inner radii, where $q/\epsilon$ (and hence $M'$ for toroidal rotation) is high. Anticipating a QuaLiKiz overprediction of $E\times{B}$ suppression of turbulence in this regime, we have implemented an optional setting to neglect the impact of rotation at inner radii in integrated modelling applications. When this assumption is applied, then in the dispersion relation solver, rotation effects are only included in a reduced $\rho_{norm}>0.4$ zone, with a linear reduction in impact from $\rho_{norm}=0.6\rightarrow0.4$. For the quasilinear flux integrals, the full rotation eigenfunction calculation is always included at all radii. Therefore, for $\rho_{norm}<0.4$, momentum transport is still predicted (due to symmetry breaking) even if rotation was not included in the growth rate spectrum calculation. We note that further motivation for the $\rho_{norm}>0.4$ assumption is provided by nonlinear electromagnetic simulations, showing that turbulence suppression by $E\times{B}$ shear is significantly weakened at high $\beta$~\cite{citr15a}, which is more likely at inner radii. 

Further caveats are as follows. The QuaLiKiz rotation model assumes a small Mach number ordering. This ordering is broken for heavy impurities, due to the higher Mach number. Thus, we set $M=0$ for all impurities with $A>20.5$ (heavier than Ne). $M'$ is not modified, since the ordering assumption for $M'$ is not necessarily broken at high $M$. Finally, due to observed spurious localized particle transport peaking when applying the QuaLiKiz rotodiffusion model, we suppress rotodiffusion from particle flux outputs. The provenance of this behaviour will be investigated in future revisions of the QuaLiKiz model. For main ion transport, rotodiffusion should be an insignificant feature.

\section{Validation with integrated modelling of JET discharges}
\label{sec:section4}
We present QuaLiKiz validation within the JETTO-SANCO~\cite{cena88,roma14} suite of integrated modelling codes, for C-wall JET hybrid scenario 75225~\cite{hobi12}, and ILW baseline scenario 87412~\cite{rimi15}. These simulations include the first ever QuaLiKiz integrated modelling simulations with combined heat, particle and momentum transport, together with the impact of rotation. All source calculations are from PENCIL (NBI) and PION (ICRH). Neoclassical transport, resistivity and bootstrap current are calculated by NCLASS~\cite{houl97}. The current profile is prescribed from predictive poloidal flux evolution modelling, using experimentally measured kinetic profiles. In JETTO-SANCO 1~s of JET plasma simulation with QuaLikiz costs $\sim10$ hours walltime using 10 CPUs. All simulations covered both ion and electron scales with a wavenumber spectrum of $k_\theta\rho_s=[0.1, 0.175, 0.25, 0.325, 0.4, 0.5, 0.7, 1, 1.8, 3 ,9 ,15, 21, 27, 36, 45]$. 

QuaLiKiz has shifted-circular geometry, while JETTO maintains shaped flux surface geometry for all source calculations, and for setting the metrics in the transport PDEs. This necessitates certain choices and transformations when passing information between the codes. The radial coordinate chosen to calculate the input gradients into QuaLiKiz is $r=\left(R_{max}-R_{min}\right)/2$, where $R_{max}$ is the maximum radius of a flux surface, and $R_{min}$ the minimum radius. The QuaLiKiz output fluxes (per $m^2$) are (circular) flux-surface-averaged, and the full flux surface area is used to calculate the total transport rates within JETTO. The effective diffusivities and heat conductivities calculated from these fluxes within QuaLiKiz are transformed into JETTO coordinates, using general geometry information from JETTO. For example for particle diffusivities, $D_{JETTO}={c_D}D_{QuaLiKiz}$, where:
\begin{equation}
c_D = \frac{\frac{dn}{dr}\langle|\nabla\rho|\rangle}{ \frac{dn}{d\rho}\langle|\nabla\rho|^2\rangle}
\end{equation}
In practice, $c_D<\sim1.2$. 

For evaluating the accuracy of the QuaLiKiz predictions, we apply a simple standard deviation figure of merit~\cite{ITER,holl16}, where, for a quantity $f$:
\begin{dmath}
	\sigma=\sqrt{\int_0^{\rho_{BC}}dx\left(f_{sim}-f_{exp}\right)^2}/\sqrt{\int_0^{\rho_{BC}}dxf^2_{exp}}
\end{dmath}
Where $f_{sim}$ is the simulated quantity, and $f_{exp}$ is the measured quantity.
We do not take into account the multiple sources of potential systematic error in the modelling, such as the heating and current drive models, $Z_{eff}$ measurements, neoclassical model, and systematic kinetic profile measurement errors. This is out of the scope of the present paper. 

\subsection{JET hybrid scenario 75225}
\begin{figure}[htbp]
	\centering
	\includegraphics[scale=0.8]{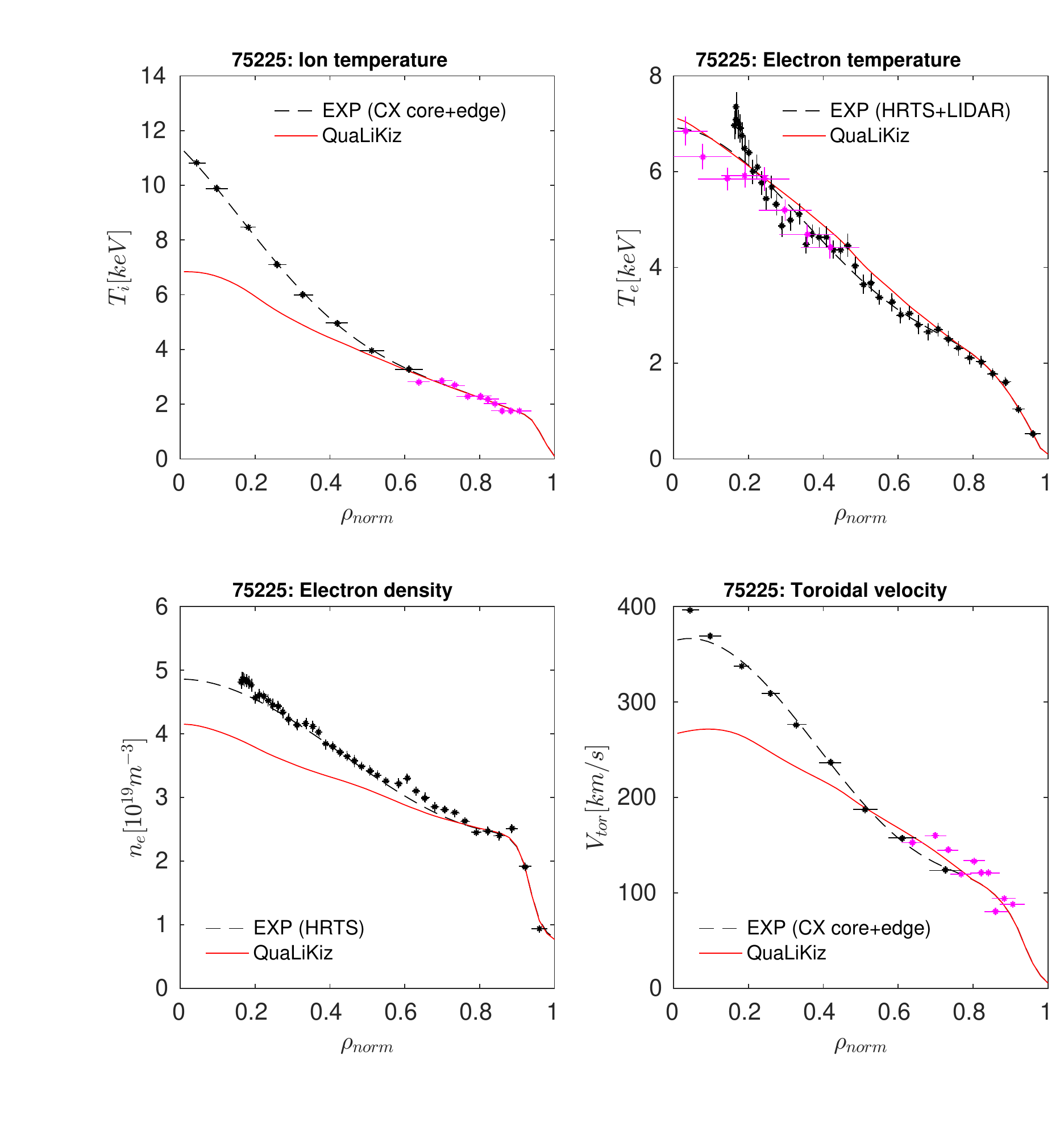}
\caption{\small Comparison between JETTO+QuaLikiz predictions for JET hybrid scenario 75225 including heat, particle and momentum transport. The QuaLiKiz predicted and measured profiles are compared, for: ion temperature (top left panel), electron temperature (top right panel), electron density (bottom left panel) and toroidal velocity (bottom right panel). The experimental data points are averaged between 6-6.5~s. The experimental error bars are statistical (and are reduced following the averaging), and do not include potential systematic errors. The QuaLiKiz boundary condition is at $\rho_{norm}=0.8$}
\label{fig:75225_1}
\end{figure}

In figure~\ref{fig:75225_1} we display a JETTO+QuaLiKiz simulation of JET C-wall hybrid discharge 75225. The measured kinetic profiles, calculated source deposition profiles, and equilibrium data from EFIT were averaged between 6-6.5s. These averaged kinetic profiles were set as the initial conditions for JETTO+QuaLiKiz modelling, carried out for 1~s, which corresponds to approximately 5 energy confinement times and sufficient for profile convergence. The QuaLiKiz predicted profiles are then compared to the averaged measured profiles. The boundary condition is at normalized toroidal flux coordinate $\rho_{norm}=0.8$. The modelling includes heat, particle, impurity, and momentum transport simultaneously. Rotation effects are always included in the $\rho_{norm}>0.4$ zone, unless otherwise specified. The C-impurity transport and radiation is evolved separately within SANCO, with transport coefficients passed from QuaLiKiz. The initial $C$ condition is a flat profile, with $Z_{eff}=1.7$ as measured by Bremmstrahlung. The fits of the experimental profiles are from a combination of edge and core charge exchange for $T_i$ and $V_{tor}$, High Resolution Thompson Scattering (HRTS) and LIDAR for $T_e$, and from HRTS for $n_e$. 

\begin{table*}[htbp]
	\centering
	\caption{\footnotesize Standard deviation figures of merit for the JETTO+QuaLiKiz JET hybrid scenario 75225 simulations shown in figure~\ref{fig:75225_1}.}
	%\vspace{0.15cm}
	\tabcolsep=0.11cm
	\scalebox{1.0}{\begin{tabular}{c|c|c|c|c}
	\label{tab:75225_1}
	Zone & $\sigma_{T_i}$ & $\sigma_{T_e}$ & $\sigma_{n_e}$ & $\sigma_{Vtor}$ \\
	\hline
	$\rho_{norm}=[0.4-0.8]$ & 7.9\% & 8.3\% & 7.7\% & 7.3\% \\
	$\rho_{norm}=[0.1-0.8]$ & 24.1\% & 5.0\% & 13.2\% & 18.8\% \\
	\end{tabular}}
\end{table*}

The simulation figures of merit per transport channel are summarized in table~\ref{tab:75225_1}. When restricting the comparison to $\rho_{norm}>0.4$,  multi-channel agreement is achieved at a $\sim$10\% level for each channel. However, at $\rho_{norm}<0.4$, QuaLiKiz underpredicts the value of $T_i$. This is expected: we did not include fast ions in the simulation, and furthermore QuaLiKiz is an electrostatic code and does not include nonlinearly enhanced electromagnetic stabilization of ITG in the saturation rule. Fast-ion-enhanced electromagnetic stabilization of ITG turbulence is an important effect for high performance hybrid scenarios at inner radii~\cite{citr13,chal15,garc15,doer16,brav16}. Pertinently, nonlinear simulations have shown electromagnetic stabilization to be a critical effect in inner radii for this specific discharge~\cite{citr15a}. 

\begin{figure}[htbp]
	\centering
	\includegraphics[scale=0.6]{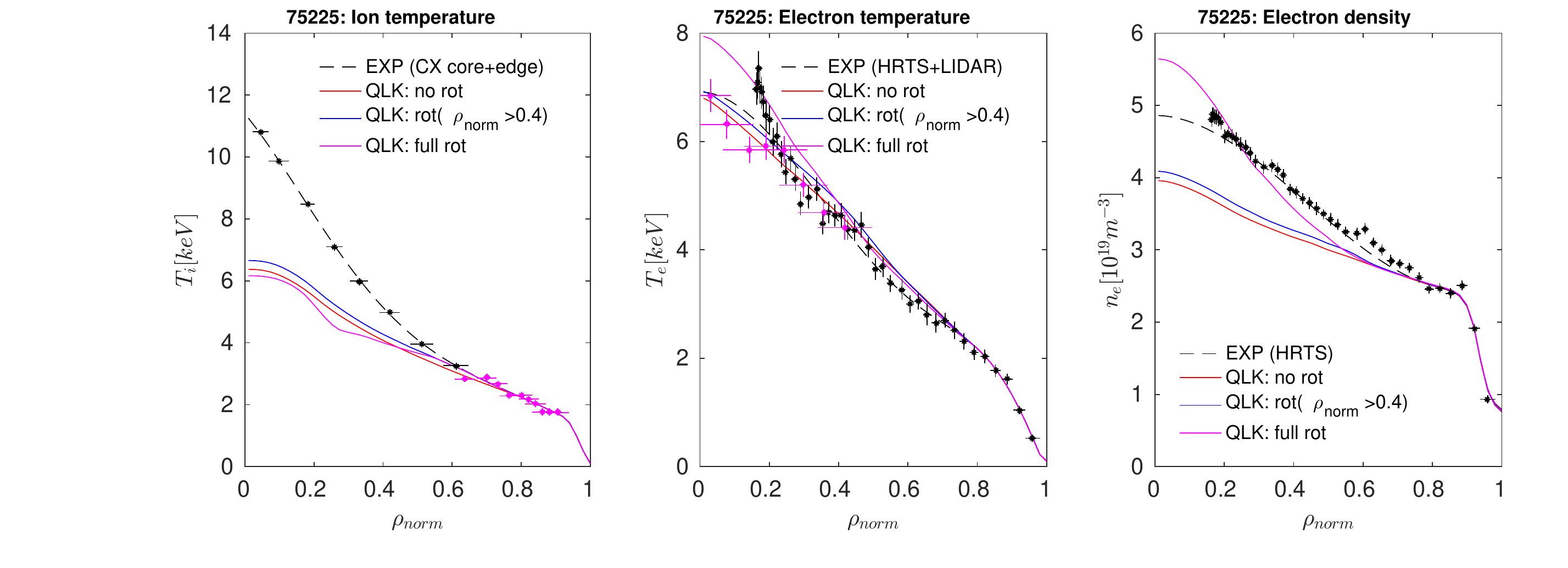}
	\caption{\small Rotation sensitivity test. Comparison between JETTO+QuaLikiz predictions for JET hybrid scenario 75225 including heat and particle transport. The QuaLiKiz predicted and measured profiles are compared, for: ion temperature (left panel), electron temperature (centre panel), electron density (right panel). 3 separate QuaLiKiz simulations are shown, with no rotation effects (red curve), with the default rotation effects in the dispersion relation limited to $\rho_{norm}>0.4$ (blue curve), and when including rotation throughout the entire profile (magenta curve). The QuaLiKiz boundary condition is at $\rho_{norm}=0.8$}
	\label{fig:75225_2}
\end{figure}

The impact of the various rotation assumptions are shown in figure~\ref{fig:75225_2}, with the figures of merit listed in table~\ref{tab:75225_2}. These simulations were carried out with heat and particle transport only. The inclusion of rotation clearly improves agreement with experiment for $\rho>0.4$. However, when including rotation throughout the full radius, then the $T_e$ and $n_e$ profiles peak at $\rho<0.4$ with gradients significantly (particularly for $T_e$) in disagreement with the measured profiles.

\begin{table*}[htbp]
	\centering
	\caption{\footnotesize Standard deviation figures of merit for the JETTO+QuaLiKiz JET hybrid scenario 75225 simulations shown in figure~\ref{fig:75225_2}. All cases are calculated for the range $\rho_{norm}=[0.1-0.8]$}
	%\vspace{0.15cm}
	\tabcolsep=0.11cm
	\scalebox{1.0}{\begin{tabular}{c|c|c|c}
	\label{tab:75225_2}
	Rotation assumption & $\sigma_{T_i}$ & $\sigma_{T_e}$ & $\sigma_{n_e}$ \\
	\hline
	No rotation & 29.3\% & 5.1\% & 16.8\%  \\
	$\rho>0.4$  & 26.2\% & 5.1\% & 14.5\% \\
	Full radius & 31.8\% & 8.3\% & 7.0\% \\
	\end{tabular}}
\end{table*}

\begin{figure}[htbp]
	\centering
	\includegraphics[scale=0.8]{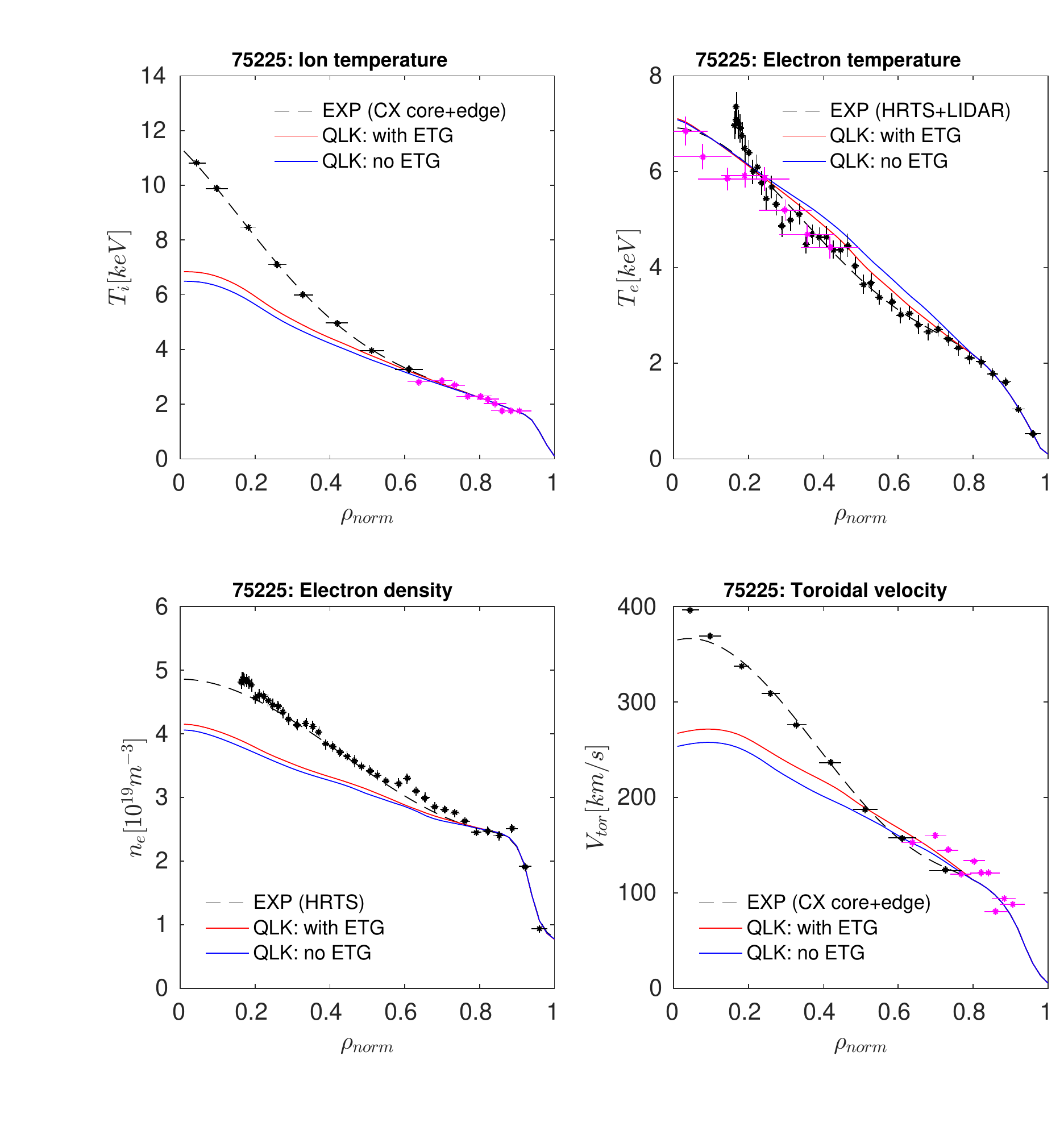}
\caption{\small ETG sensitivity test. Comparison between JETTO+QuaLikiz predictions for JET hybrid scenario 75225 including heat, particle and momentum transport. The QuaLiKiz predicted and measured profiles are compared, for: ion temperature (upper left panel), electron temperature (upper right panel), electron density (lower left panel) and toroidal velocity (lower right panel). 2 separate QuaLiKiz simulations are shown, including ETG effects (red curve), and without ETG effects (blue curve). The QuaLiKiz boundary condition is at $\rho_{norm}=0.8$}
\label{fig:75225_3}
\end{figure}

The impact of including ETG scales is shown in figure~\ref{fig:75225_3}. The figures of merit are listed in table~\ref{tab:75225_3}. These simulations include heat, particle and momentum transport. Ignoring the ETG scales leads to an overprediction of the $T_e$ gradients at $\rho_{norm}>0.5$. This is consistent with previous observations of an electron heat flux shortfall at outer radii in DIIID modelling, using a previous version of TGLF~\cite{stae07,kins08} with only $\sim10\%$ electron heat flux from ETG scales~\cite{kins15}. Interestingly, excluding ETG scales leads to cross-channel transport phenomena, where $V_{tor}$ is also significantly modified, likely due to $\nabla{T_e}$ or $T_i/T_e$ dependencies in the momentum transport coefficients.

\begin{table*}[htbp]
	\centering
	\caption{\footnotesize Standard deviation figures of merit for the JETTO+QuaLiKiz JET hybrid scenario 75225 simulations shown in figure~\ref{fig:75225_3}. Figures of merit calculated in the ranges $\rho_{norm}=[0.1-0.8]$ and $\rho_{norm}=[0.4-0.8]$ are both displayed}
		%\vspace{0.15cm}
	\tabcolsep=0.11cm
	\scalebox{1.0}{\begin{tabular}{c|c|c|c|c|c}
	\label{tab:75225_3}
	Zone & ETG assumption & $\sigma_{T_i}$ & $\sigma_{T_e}$ & $\sigma_{n_e}$ & $\sigma_{Vtor}$ \\
	\hline
$\rho=[0.4-0.8]$ & With ETG & 7.9\% & 8.3\% & 7.7\% & 7.3\% \\
$\rho=[0.4-0.8]$ & No ETG   & 11.0\% & 14.1\% & 9.0\% & 10.3\% \\
	\hline
$\rho=[0.1-0.8]$ & With ETG & 24.1\% & 5.0\% & 13.2\% & 18.8\% \\
$\rho=[0.1-0.8]$ & No ETG   & 27.5\% & 8.4\% & 15.0\% & 22.5\% \\
	\end{tabular}}
\end{table*}

We summarize the 75225 simulation results. Encouraging agreement in heat, particle, and momentum transport has been achieved. The most outstanding disagreement remains $T_i$ underprediction in inner radii, consistent with missing electromagnetic physics in QuaLiKiz. Both the inclusion of rotation at $\rho_{norm}>0.4$ (where the rotation model is valid), and ETG scales, improves agreement with experiment.

\subsection{JET baseline scenario 87412}
\subsubsection{Heat, particle and momentum tranport}
\begin{figure}[htbp]
	\centering
	\includegraphics[scale=0.8]{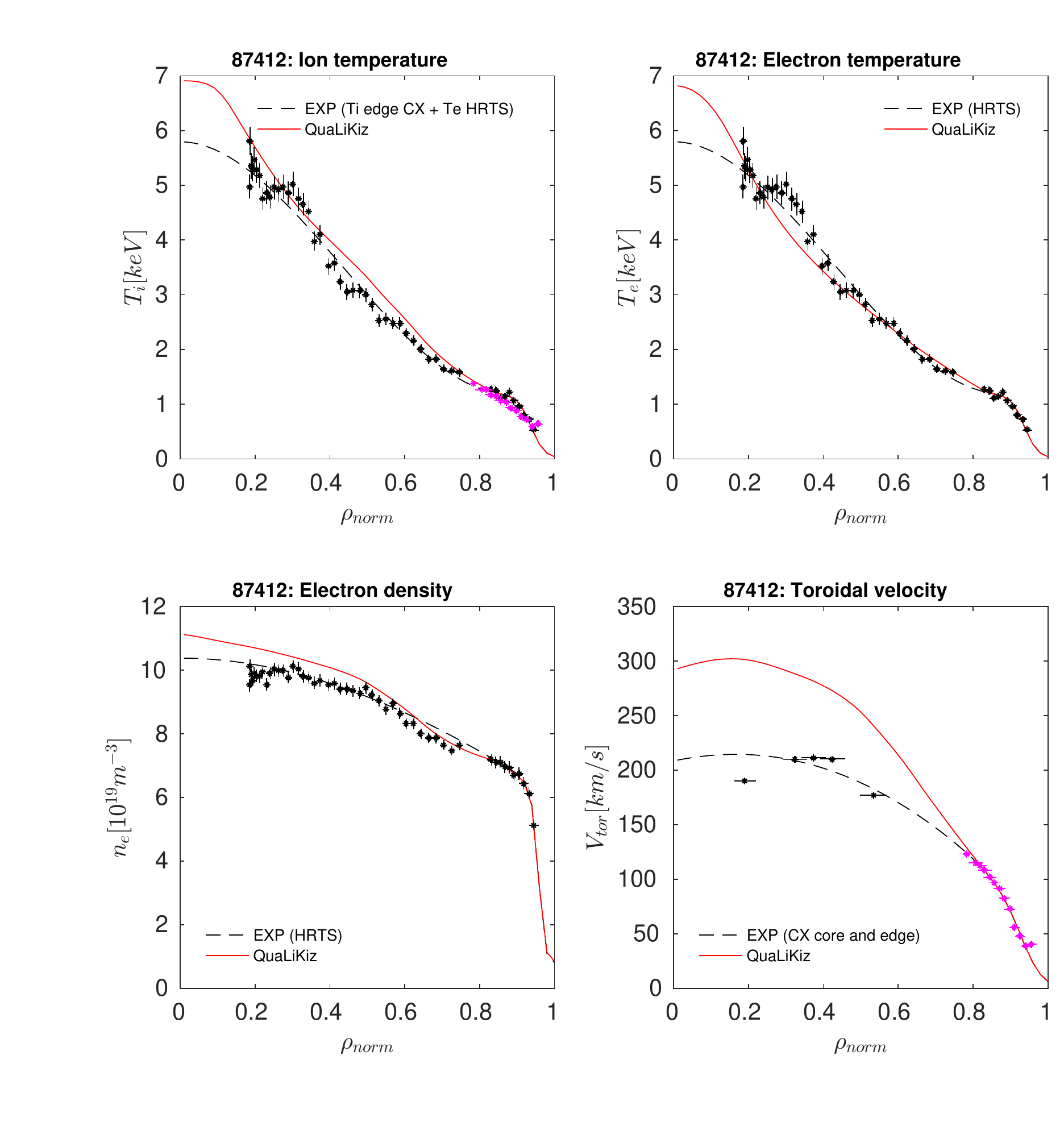}
	\caption{ \small Comparison between JETTO+QuaLiKiz predictions for JET baseline scenario 87412 including heat, particle and momentum transport. The QuaLiKiz predicted and measured profiles are compared, for: ion temperature (top left panel), electron temperature (top right panel), electron density (bottom left panel) and toroidal velocity (bottom right panel). The experimental data points are averaged between 10-10.5~s. The experimental error bars are statistical (and are reduced following the averaging), and do not include potential systematic errors. The QuaLiKiz boundary condition is at $\rho_{norm}=0.85$}
	\label{fig:87412_1}
\end{figure}

A similar validation exercise was carried out for JET ILW baseline discharge 87412. Measured kinetic profiles, calculated source deposition profiles, and equilibrium data from EFIT were averaged between 10-10.5s. These averaged kinetic profiles were set as the initial conditions for JETTO+QuaLiKiz modelling, carried out for 1~s, approximately 4 energy confinement times. The QuaLiKiz predicted profiles are then compared to the averaged measured profiles. The source profiles were prescribed constant during the simulation. The boundary condition is at normalized toroidal flux coordinate $\rho_{norm}=0.85$. Heat, particle, and momentum transport were included. Rotation effects are always included in the $\rho>0.4$ zone, unless otherwise specified. Due to significant uncertainties in the core CX measurements for this discharge, we assume $T_i=T_e$ in the core, an assumption strengthened by the high density typical of baseline scenarios~\cite{beur13}. For the edge parameters, this is validated by comparing the $T_e$ HRTS and $T_i$ measurements from edge CX, as seen in the top left panel in figure~\ref{fig:87412_1}. The Be-impurity transport and radiation is evolved separately within SANCO, with transport coefficients passed from QuaLiKiz. The initial $Be$ condition is a flat profile, with $Z_{eff}=1.2$ as measured by Bremmstrahlung. No W was included in these simulations, which concentrate on a phase without W-accumulation. 

The full comparison is shown in figure \ref{fig:87412_1}. We summarize the prediction figures of merit in table~\ref{tab:87412_1}, We did not include the (infrequent) sawteeth in the simulations, which may lead to spurious inner core peaking in the predictions. We thus limit the figure of merit radial range to $\rho_{norm}=[0.2-0.85]$. Note the excellent agreement in the heat and particle channels. As opposed to the hybrid scenario 75225, no $T_i$ underprediction is seen here. This is consistent with electromagnetic effects being of less importance in the lower $\beta$ baseline scenarios. The overprediction of $V_{tor}$ may also be affected by NTV torque from magnetic islands. 3/2 and 4/3 modes are present during the studied time window. Their impact was not taken into account. Nevertheless, all profiles are in a stationary state by the end of the simulation. This is not trivial considering that a potential feedback mechanism could occur when including momentum transport. Overpredicted $E{\times}B$ shear may stabilize the turbulence, leading to reduced momentum transport and a further buildup of ExB shear, forming an Internal Transport Barrier. This did not occur for the two discharges simulated here.

\begin{table*}[htbp]
	\centering
	\caption{\footnotesize Standard deviation figures of merit for the JETTO+QuaLiKiz baseline 87412 simulation shown in figure~\ref{fig:87412_1}. The figures of merit are calculated in the ranges $\rho_{norm}=[0.2-0.85]$}
	%\vspace{0.15cm}
	\tabcolsep=0.11cm
	\scalebox{1.0}{\begin{tabular}{c|c|c|c}
	\label{tab:87412_1}
	$\sigma_{T_i}$ & $\sigma_{T_e}$ & $\sigma_{n_e}$ & $\sigma_{Vtor}$ \\
	\hline
	7.9\% & 6.6\%  & 4.2\% & 33.1\%
	\end{tabular}}
\end{table*}

\begin{figure}[htbp]
	\centering
	\includegraphics[scale=0.8]{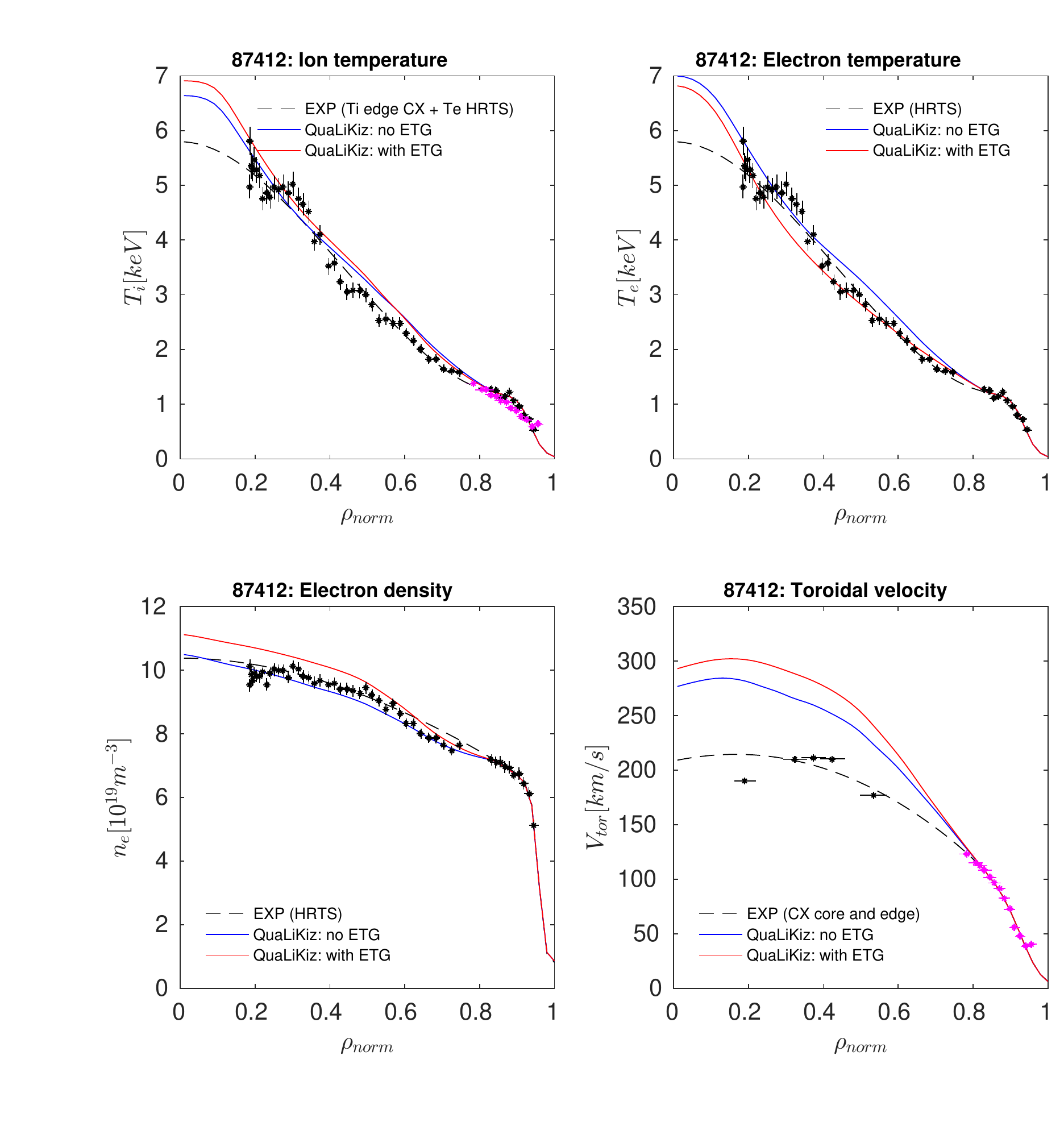}
	\caption{ \small Comparison between JETTO+QuaLiKiz predictions for JET baseline scenario 87412 including heat, particle and momentum transport. The QuaLiKiz predicted and measured profiles are compared, for: ion temperature (top left panel), electron temperature (top right panel), electron density (bottom left panel) and toroidal velocity (bottom right panel). The QuaLiKiz boundary condition is at $\rho_{norm}=0.85$. Two QuaLiKiz cases are compared, neglecting (blue curve) and including (red curve) ETG scales. The results are shown following 1~s simulation time.}
	\label{fig:87412_3}
\end{figure}

The impact of neglecting the ETG scales is shown in figure~\ref{fig:87412_3}, and the figures of merit are summarized in table~\ref{tab:87412_2}. Neglecting ETG scales leads to a deterioration of $T_e$ agreement in the outer-half radius, again underlying the importance of ETG scales. 

\begin{table*}[htbp]
	\centering
	\caption{\footnotesize Standard deviation figures of merit for the JETTO+QuaLiKiz baseline 87412 simulations shown in figure~\ref{fig:87412_3}. The figures of merit are calculated in the ranges $\rho_{norm}=[0.2-0.85]$. Results are shown following 1~s of simulation time, including and neglecting ETG scales}
	%\vspace{0.15cm}
	\tabcolsep=0.11cm
	\scalebox{1.0}{\begin{tabular}{c|c|c|c|c}
	\label{tab:87412_2}
	ETG assumption & $\sigma_{T_i}$ & $\sigma_{T_e}$ & $\sigma_{n_e}$ & $\sigma_{Vtor}$ \\
	\hline
	With ETG & 7.9\% & 6.6\%  & 4.2\% & 33.1\% \\
	No ETG   & 6.1\% & 7.3\% & 3.2\% & 24.1\% \\
	\end{tabular}}
\end{table*}

\subsubsection{Heat and particle tranport only}
We examine the sensitivity of the QuaLiKiz transport predictions for JET 87412 to rotation. Since momentum transport is identically zero with the QuaLiKiz rotation model turned off (no symmetry breaking), this sensitivity scan was carried out with heat and particle transport only. 

\begin{figure}[htbp]
	\centering
	\includegraphics[scale=0.6]{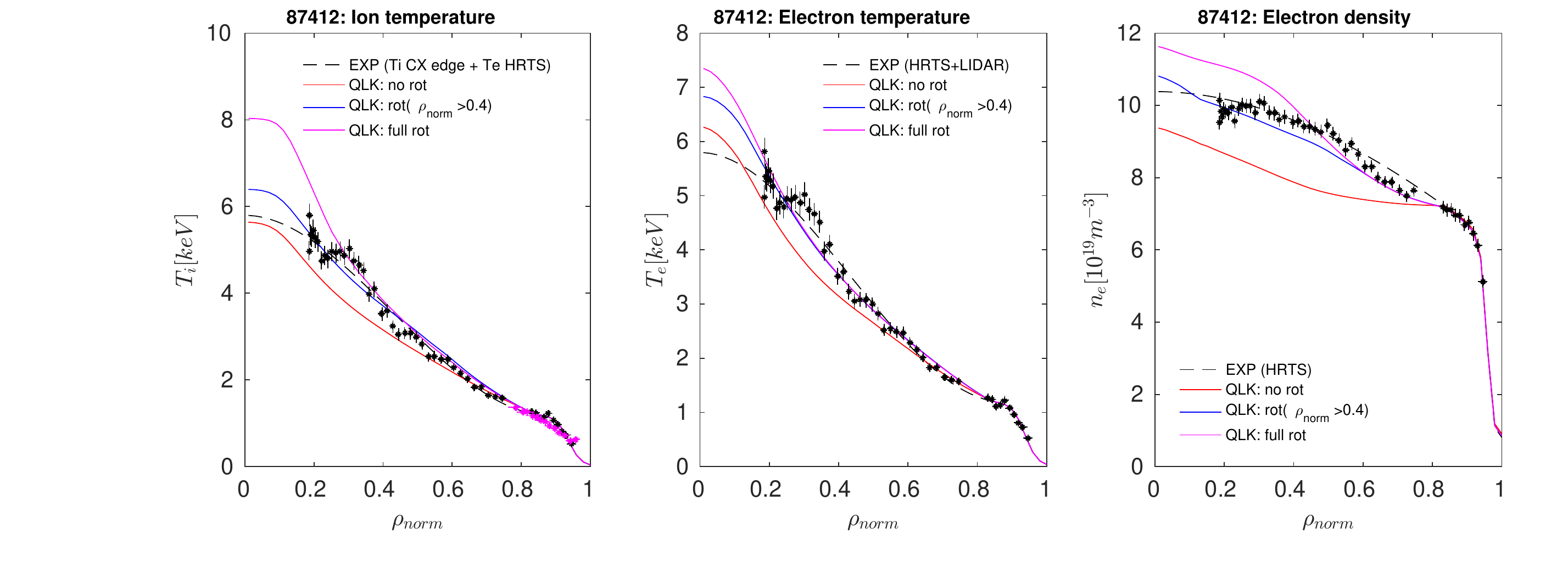}
	\caption{ \small Rotation sensitivity test. Comparison between JETTO+QuaLikiz predictions for JET baseline scenario 87412 including heat and particle transport. The QuaLiKiz predicted and measured profiles are compared, for: ion temperature (left panel), electron temperature (centre panel), electron density (right panel). 2 separate QuaLiKiz simulations are shown, with no rotation effects (red curve), and with the default rotation effects in the dispersion relation limited to $\rho_{norm}>0.4$ (blue curve), and full rotation effects throughout the entire radius (magenta curve). The QuaLiKiz boundary condition is at $\rho_{norm}=0.85$}
	\label{fig:87412_4}
\end{figure}

In figure~\ref{fig:87412_4}, we present the rotation sensitivity test. The figures of merit are listed in table~\ref{tab:87412_4}. The inclusion of rotation has a striking impact on all profiles, significantly improving agreement when rotation is included at $\rho_{norm}>0.4$. However, when including rotation throughout the full radius, then the $T_i$ and $n_e$ profiles peak at $\rho<0.4$ with gradients significantly exceeding those of the measured profiles. These observations support the default choice of only including rotation effects in the dispersion relation for $\rho_{norm}>0.4$.

\begin{table*}[htbp]
	\centering
	\caption{\footnotesize Rotation sensitivity test. Standard deviation figures of merit for the JETTO+QuaLiKiz baseline 87412 simulations shown in figure~\ref{fig:87412_4}. The figures of merit are calculated in the ranges $\rho_{norm}=[0.2-0.85]$. Results are shown following 1~s of simulation time, including and neglecting rotation}
	%\vspace{0.15cm}
	\tabcolsep=0.11cm
	\scalebox{1.0}{\begin{tabular}{c|c|c|c}
			\label{tab:87412_4}
			Rotation assumption & $\sigma_{T_i}$ & $\sigma_{T_e}$ & $\sigma_{n_e}$  \\
			\hline
			No rotation & 14.7\% & 13.5\% & 14.9\% \\
			$\rho>0.4$ & 4.2\% & 4.7\%  & 4.5\% \\
			Full radius & 10.3\% & 4.9\%  & 5.9\% \\
		\end{tabular}}
	\end{table*}

\subsubsection{Time dependent simulations}
The validations discussed so far concentrated on stationary state. However, an important application for integrated modelling is profile dynamics. For example, the onset of W-accumulation is often set by a threshold in density peaking~\cite{angi14,cass15}. Profile dynamics were examined for the 87412 density rise following the LH transition. Heat and particle transport JETTO+QuaLiKiz simulations were carried out for discharge 87412, starting from an initial condition at $t_{exp}=9~s$, approximately 300ms following the LH transition.  The $T_e$, $T_i$, and $n_e$ boundary conditions at $\rho_{norm}=0.85$ evolve in time following the experimental profile fits. 

\begin{figure}[htbp]
	\centering
	\includegraphics[scale=0.6]{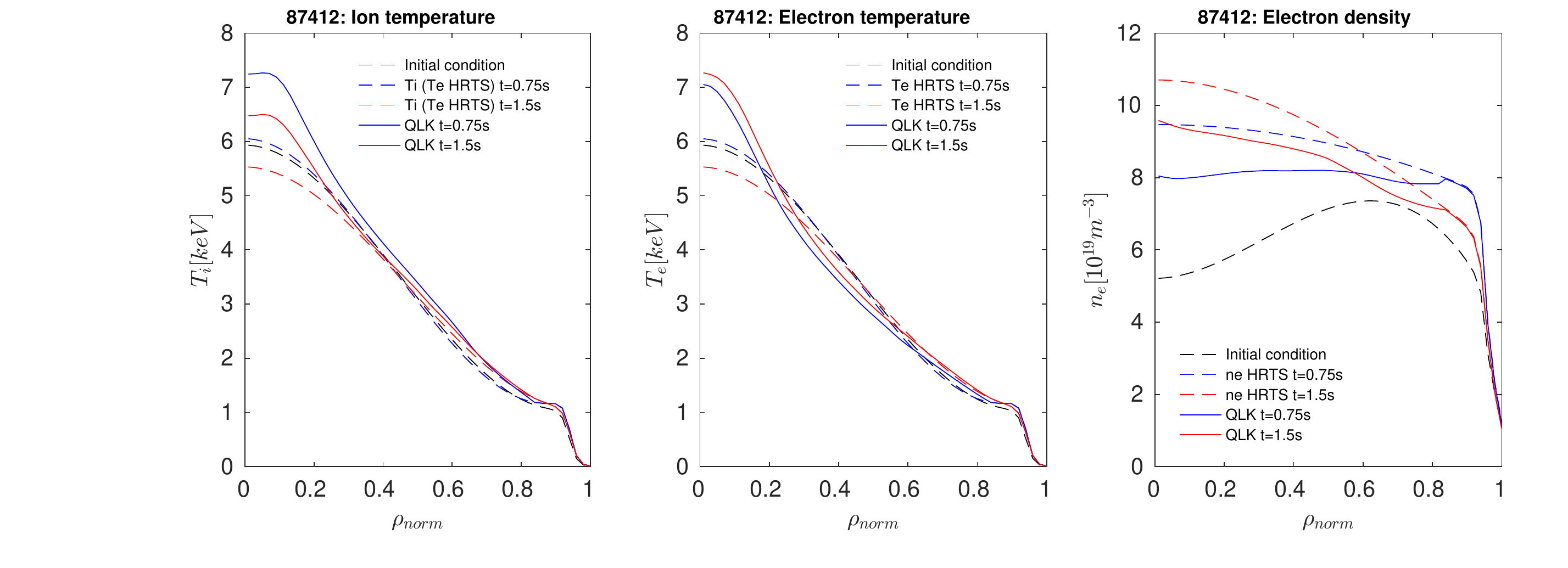}
	\caption{ \small Comparison of JETTO+QuaLikiz time dependent predictions and measured profiles for JET baseline scenario 87412, including heat and particle transport only, for: ion temperature (left panel), electron temperature (centre panel), and electron density (right panel). The initial condition at $t_{sim}=0$ (black dashed curve) is at $t_{exp}=9~s$, approximately 300ms following the LH transition. The measured (dashed) and QuaLiKiz predictions (solid curves) are compared at t=0.75s (blue curves) and t=1s (red curves)}
	\label{fig:87412_7}
\end{figure}

The results are shown in figure~\ref{fig:87412_7}. The particle confinement time is significantly longer than the heat confinement time, as evidenced from the relatively small variation in $T_i$ and $T_e$ profiles over the 1.5~s of simulation time. However, the initial $n_e$ profile is hollow, and slowly evolves towards a peaked profile. This essential behaviour is reproduced by QuaLiKiz, similar to previous work within CRONOS~\cite{arta10,baio15a}. However, the QuaLiKiz $n_e$ evolution is slower than the measured $n_e$ evolution for this case, with the predicted degree of peaking less than observed.

\begin{figure}[htbp]
	\centering
	\includegraphics[scale=0.8]{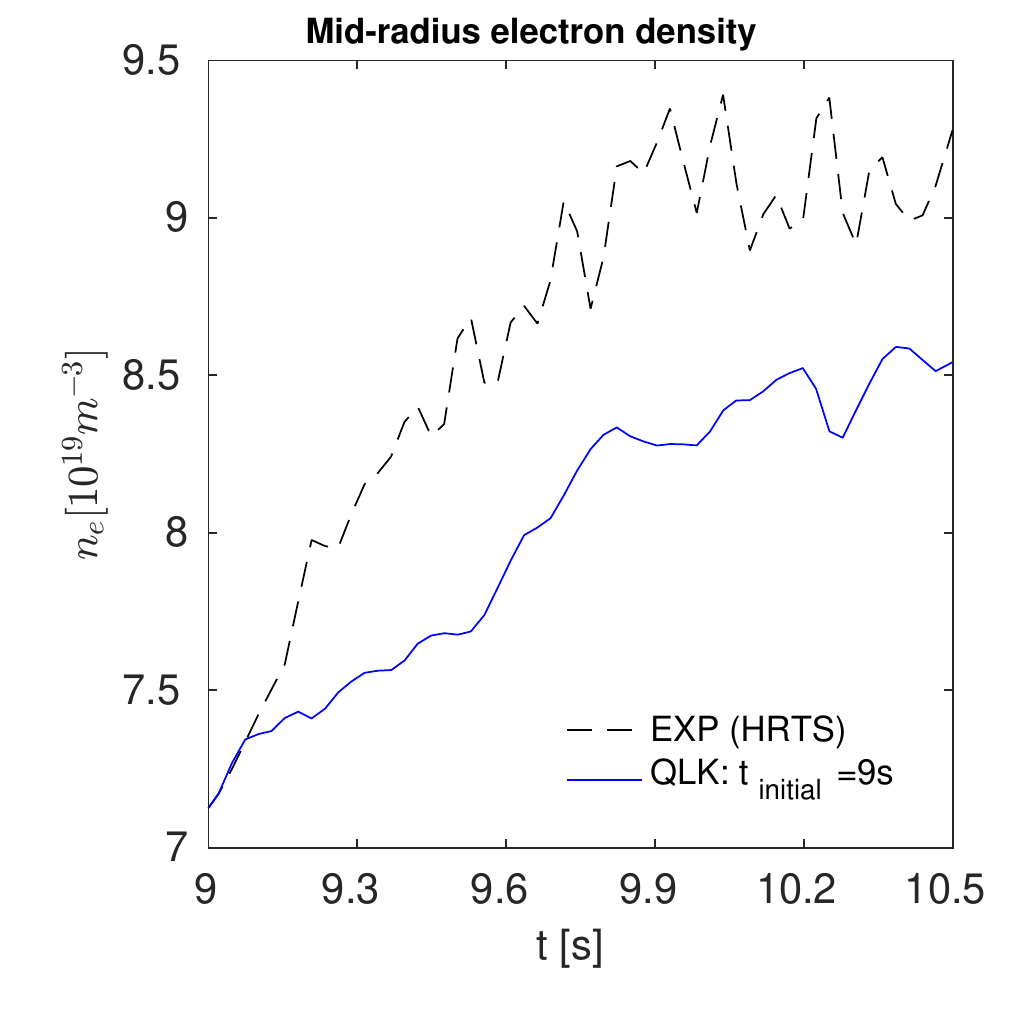}
	\caption{ \small Comparison of time-dependent JETTO+QuaLikiz and measured time evolution of the electron density at mid-radius for JET baseline scenario 87412. The measured (from HRTS) curve is black-dashed. The QuaLiKiz predictions are shown starting from t=9~s (blue curve).}
	\label{fig:87412_8}
\end{figure}

Further analysis shows that the $n_e$ discrepancy is limited to the time between 9.1-9.6~s. This is shown in figure~\ref{fig:87412_8}, which displays $n_e$ time-traces at mid-radius, from measurements and the QuaLiKiz prediction starting at $t_{initial}=9~s$. The significant observed increase in $n_e$ between t=9.1-9.6~s in not captured by QuaLikiz. Outside of this range, the experimental trend is well followed. We did not include the (infrequent) sawteeth in the simulations, which are observed for this discharge. This may play a role in accelerating the filling in of the hollow density profile. Future work will include such MHD effects. Nevertheless, the essential behaviour of the inversion from hollow to peaked profiles is captured.

\section{Conclusions and outlook}
The first-principle-based quasilinear gyrokinetic transport code QuaLiKiz is numerically optimized. Eigenvalue calculations take $\sim1$~CPUs, and flux calculations take $\sim10$~CPUs. This demands approximately 100~CPUh computation time for 1~s of JET plasma evolution within the JETTO integrated modelling suite, comparable to the TGLF quasilinear transport model. The physics content of the code has been extended and validated, including poloidal asymmetry effects important for heavy impurity transport, and a recalibrated ETG model. Successful validation was carried out within the JETTO modelling suite, for JET hybrid scenario 75225 and baseline scenario 87412, including the first QuaLiKiz integrated modelling simulations with combined heat, particle and momentum transport. Both the impact of rotation and ETG scales was shown to be important for improving agreement with experiments. 

QuaLiKiz is now ready for extensive integrated modelling applications, including for W-transport, which is ongoing work. Such multi-channel integrated modelling is a powerful tool for self-consistent kinetic profile prediction, including the dynamics of interacting turbulent transport channels (heat, particle and momentum) as well as setting the background profiles for neoclassical impurity transport. Future work on the QuaLiKiz physics model will concentrate on alleviating the observed inconsistencies in $\alpha$-stabilization and parallel velocity gradient destabilization, as seen in comparisons with linear gyrokinetic modelling using more complete models such as \textsc{Gene} and GKW. Extension to shaped geometries will be explored. Furthermore, a model for nonlinear electromagnetic stabilisation must be incorporated in the ion-scale turbulence saturation rule. Further computation acceleration aimed at extensive scenario optimization and realtime applications is being carried out by emulation of the QuaLiKiz model using neural network regression~\cite{citr15b}.

\textit{Acknowledgements}.--
This work is part of the research programme `Fellowships for Young Energy Scientists' (YES!) of the Foundation for Fundamental Research on Matter (FOM), which is financially supported by the Netherlands Organisation for Scientific Research (NWO). DIFFER is a partner in the Trilateral Euregio Cluster TEC. The work has been carried out within the framework of the EUROfusion Consortium and has received funding from the Euratom research and training programme 2014-2018 under grant agreement No 633053. The views and opinions expressed herein do not necessarily reflect those of the European Commission. 

%\section*{References}
\singlespacing
\bibliographystyle{unsrt}
\bibliography{qlk2016bib}

\begin{appendix}
\section{Numerical optimization of QuaLiKiz dispersion relation}
\label{app:opt}
The optimization has 3 components: tailoring the contour paths in the $D(\omega)$ root finding algorithm, improved numerical techniques for integral functions within $D(\omega)$, and taking advantage of the existence of previous solutions from previous timesteps within integrated modelling applications.
\subsection{Contour path optimization}
The root finding algorithm in the QuaLiKiz eigenvalue solver follows the Davies method~\cite{davi86}. This involves calculating contours of $D(\omega)$ in the complex plane and invoking the argument principle to identify and set the initial guess for any roots found within each given contour. A Newton method is then applied to converge on the root.

Optimizing the precise contour paths is key to reducing the computation time, by minimizing the number of individual $D(\omega)$ evaluations carried out. This optimization involves: setting the boundaries of the eigenvalue search region in the complex plane, splitting the search region into a number of separate contours, setting the degree of contour overlap, and parametrizing the contour shape. Previously, the contour paths were optimized for robustness and not for speed in integrated modelling applications, meaning extensive limits and significant contour overlap was in place. We now summarize the newly applied contour path choices, which significantly speed up the solution time while maintaining a high level of robustness.
\subsubsection*{Search region limits:}
Only solutions in the upper (unstable) half of the complex plane are sought out. The ceiling of the contours in the imaginary plane is set by calculating simple analytical estimates of both slab and interchange modes, calculated by a simplified high $\omega$ expansion of the dispersion relation~\cite{bourphd}. The actual ceiling then corresponds to $max(\gamma_{interchange},\gamma_{slab})/1.5$. The analytic growth rate tends to significantly overestimate the kinetic growth rate, hence the reduction of factor $1.5$ (decided upon following a series of trials), which constrains the contour to the zone where the instabilities are more likely to be found (increasing robustness), while simultaneously not risking missing the growth rate by undershooting the contour ceiling.

The limit of the contour search region on the real axis, both in the negative (ion modes in the QuaLiKiz convention) and positive (electron mode) directions, is set by $|max(\omega^*_i,\omega^*_e)|/2$, where $\omega^*_{i,e}$ is the ion/electron diamagnetic frequency. The factor $1/2$, also set following a series of trials, constrains the search region to the zone most likely to contain the instabilities. For ETG scales ($k_\theta\rho_s>2$), only instabilities in the electron diamagnetic direction are considered, saving CPU time due to avoiding the calculation of unnecessary contours.
\subsubsection*{Contour shape parameterization:}
The Davies algorithm is initially conceived for circular contours. We transform the circular contours onto paths in the upper half of the complex plane within our search region limits. To minimize the probability of skipping roots, the contours should maximize the total encompassed area in the complex plane. To achieve this, a mapping of contours onto squircle paths is carried out~\cite{fong15}. This is an improvement from the previous elliptical transformations in QuaLiKiz. We apply a squircular mapping from the normalized circle radius at 0.975, which avoids discontinuities at the squircle corners which otherwise may lead to difficulty in numerical convergence. The squircles are then translated and stretched into a more rectangular shape, to the desired form and location on the complex plane.
\subsubsection*{Contour splitting and overlap:}
In total, 2 squircle contours are employed on both the positive and negative sides of the real axis. In the central region spanning the real axis, a more narrow contour is applied, with a 10\% overlap with each of the neighbouring contours. This central contour increases the algorithm robustness, since numerical evaluation of $D(\omega)$ is more difficult near the real axis, while most instabilities do not have frequencies with $\omega_r\sim0$.  
\\

\begin{figure}[htbp]
	\centering
	\includegraphics[scale=0.9]{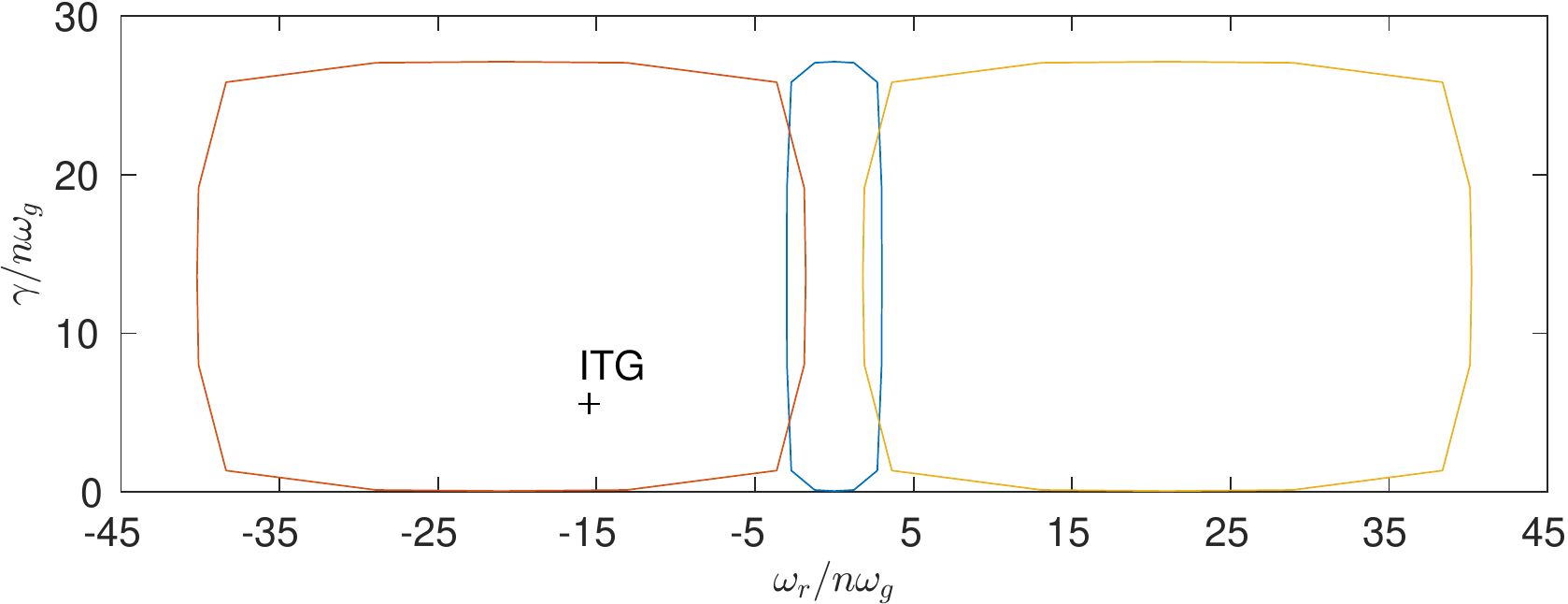}
	\caption{\footnotesize Contours applied in dispersion relation solution for collisionless GA-STD case at $k_\theta\rho_s=0.5$. The location of the ITG mode within the contour is displayed.}
	\label{fig:contours}
\end{figure}

An example of the contour paths are shown in figure~\ref{fig:contours}, for the GA Standard Case parameter set~\cite{kins05} at $k_\theta\rho_s=0.5$. An ITG mode is present, indicated in the figure. Each separate contour is signified by a different colour. The frequencies and growth rates are in units of $n\omega_g\equiv\frac{k_\theta\bar{T}}{eRB}$, where $\bar{T}\equiv1~keV$. 

The optimized contour choices led to a significant speed-up of factor $\sim$8 compared to the previous QuaLiKiz version.

\subsection{Improved numerical methods for integral functions}
The plasma dispersion function is extensively called during the evaluation of $D(\omega)$, taking up a significant fraction of the computation time. This calculation time was reduced by switching from the standard ``WofZ'' method~\cite{gaut69} to the faster Weideman algorithm~\cite{weid94}. The Weideman method is based on a rational expansion whose coefficients are pre-calculated and tabulated. This speed-up does not lead to a reduction in accuracy. A comparison between the WofZ and Weideman methods for plasma dispersion function calculations resulted in relative agreements within 0.005\%, with these maximum differences only reached in the vicinity of the function zeros~\cite{gurc14}. In the same reference, local toroidal ITG calculations using either Weideman and WofZ resulted in negligible relative growth rate and frequencies differences of $\sim10^{-7}$. Considering that the plasma dispersion relations are essentially identical between WofZ and Weideman, it is reasonable to assume that similar results are achieved for TEM and ETG calculations. In QuaLiKiz, a factor $2$ speed-up in the eigenvalue solver was achieved when switching from WofZ to Weideman.

Following a profiling of the code performance, a bottleneck due to elliptic integral evaluations was found. Hence, further speed-up was achieved by switching to a faster algorithm for elliptic integral calculations, from the Carlson method~\cite{carl95} to the T. Fukushima method~\cite{fuku15}. Elliptic integrals emerge in QuaLiKiz due to bounce averaging of trapped particles. The algorithm improvement was of particular importance for trapped electrons in the presence of collisions. This is because the trapped electron energy integral then no longer takes a form that can be reduced to a plasma dispersion function. The velocity space integral must then be calculated numerically, involving multiple calls of the elliptic integrals. Switching to the T. Fukushima algorithm led to a a further factor $1.5$ speed-up in the QuaLiKiz eigenvalue solver calculation.

\subsection{Code acceleration within integrated modelling applications}
A significant further speed-up in computation time can be gained specifically in integrated modelling applications. This is based on the timescale separation between the ${\Delta}t$ timestep in the transport PDE solver, and the energy confinement and current diffusion times. While the transport coefficients are calculated (i.e. a QuaLiKiz call) every ${\Delta}t$ ($\sim1~ms$ in a JET simulation), only relatively small modifications of the input parameters (kinetic profiles and magnetic equilibrium) occur within this time. Thus, instead of calculating the full contour solution at every QuaLiKiz call, the full calculation is only carried out every $n^{\mathrm{\tiny th}}$ call.
In the intermediate calls, QuaLiKiz skips the contour evaluation step, and proceeds straight to the Newton solver, using the previous eigenvalue solution from the $n-1$ step (saved within memory), as the initial guess. In this manner, the evolution of the eigenvalues are tracked on the complex plane. At each timestep, the displacement of the solution due to the evolving input parameters is small enough such that the Newton solver easily converges, based on the previous solution as the initial guess. The full contour solution is still necessary at periodic intervals, to allow the capture of new eigenvalues which may have appeared in the interim period, due to the evolving input profiles. It is not feasible in QuaLiKiz to calculate damped modes and follow them into the instability region, due to singularities in the dispersion relation integrals on the real axis which would excessively increase the computation time.

A subtle point for the intermediate steps (straight-to-Newton) must be taken into consideration if no new solution is found, e.g. if the mode is stabilized due to reduced driving gradients since the last call. In that case, while that (stable) solution is passed to the saturation rule, resulting in zero or lowered fluxes, the last \textit{unstable} solution for that wavenumber must still be kept as the initial guess for the next step. If this does not occur, then once a mode is predicted stable, it will remain stable until the next $n^{\mathrm{\tiny th}}$ call when the full contour solution is once again carried out. 

\begin{figure}[htbp]
	\centering
	\includegraphics[scale=0.9]{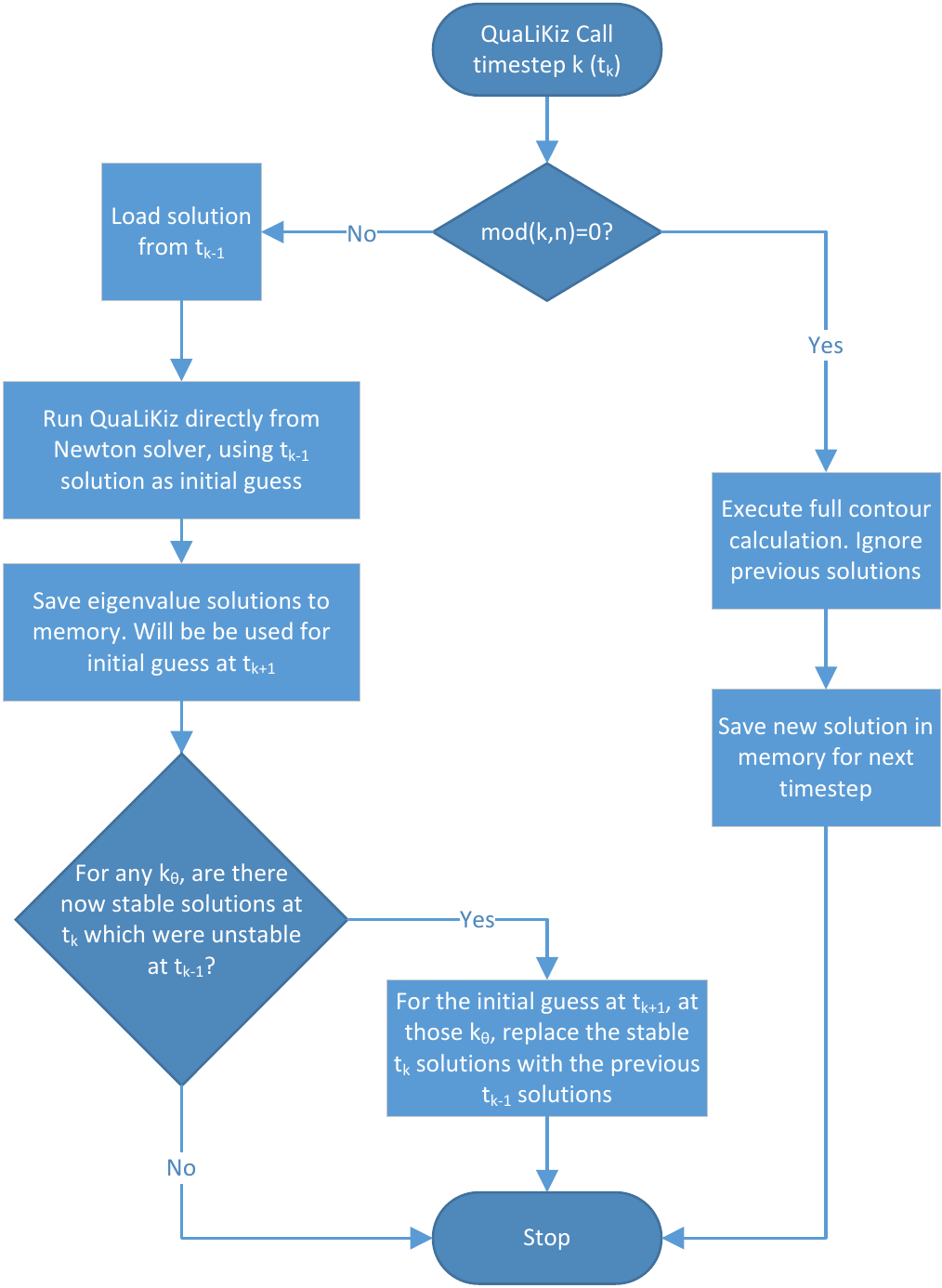}
	\caption{\footnotesize Flowchart of the straight-to-Newton procedure}
	\label{fig:flowchart}
\end{figure}

We have determined that an $n$ corresponding to $\sim10~ms$ of plasma evolution provides a reasonable balance between speed-up and robustness. For small integrated modelling timesteps, a maximum $n=20$ is set. The procedure is summarized in the flow chart in figure~\ref{fig:flowchart}. This method leads to a further factor $\sim$5 speed-up of QuaLiKiz application within an integrated modelling environment. 

\section{Improved eigenfunction solution method}
\label{app:fluid}
\subsection{Ion scales}
The starting point is the local dispersion relation:
\begin{equation}
D(\omega)=\sum_s\langle\frac{f_0q_s^2}{T_s}\left(1-J_{0s}^2\frac{\bar{\omega}-n\omega^*_s}{\bar{\omega}-k_\parallel{V_\parallel}-n\omega_d}\right)\rangle\phi(x)=0
\end{equation}
$\langle~\rangle$ signifies integration over velocity space, $s$ is a species identifier, $n=\frac{k_{\theta}r}{q}$ is the  toroidal mode number corresponding to the poloidal wavenumber $k_\theta$. $f_0$ is the background Maxwellian, $q_s$ the species charge, $T_s$ the species temperature, $J_{0s}$ the Bessel function from orbit averaging (whose arguments will be discussed below), $\bar{\omega}$ the Doppler shifted mode frequency in the rotating frame $\omega-k_\theta\gamma_Ex$ (where $x$ is the distance from the mode surface, and $\gamma_E$ the perpendicular mean flow shear), $k_\parallel$ the parallel wave number $k_\theta\frac{\shat{x}}{qR}$ (with $\shat$ the magnetic shear, $q$ the q-profile, $x$ the distance from the resonant surface, $R$ the major radius), and $V_\parallel$ the parallel velocity. $\omega*_s$ is the diamagnetic frequency:
\begin{dmath}
	\omega*_s=-\omega_{ds}\left[\rlns+\rlts\left(\xi-\frac{3}{2}-M\left(2\vbar-M\right)\right)+\rlus2\left(\vbar-M\right)\right]
\end{dmath}  
$\omega_{ds}\equiv\frac{T_sq}{q_sBrR}$. $M$ is the Mach number $U_\parallel/V_{Ts}$, $U_\parallel$ being the mean flow in the background Maxwellian, and $V_{Ts}$ the thermal velocity. $\xi=\frac{V_\parallel^2+V_\perp^2}{V_{Ts}^2}$, the normalized kinetic energy. $R/L_{T,s,u}$ are the normalized logarithmic gradients of the temperature, density, and velocities respectively. $\vbar$ is the normalized velocity (phase space coordinate) $V_\parallel/V_{Ts}$. $\omega_d$ is the magnetic drift frequency, whose expression differs for the trapped and passing species due to the trapped particle bounce averaging, as will be discussed below. $\phi(x)$ is the electrostatic potential, whose functional form we seek as the eigenfunction solution of $D(\omega)=0$, together with the eigenvalue $\omega$.

The key point in the approximated eigenfunction solution is a large $\ombar$ expansion, where $\ombar/\omega_d,\ombar/k_\parallel{V}_\parallel\ll1$. The resonant denominators are expanded to second order, leading to: 
\begin{dmath}
	\label{eq:disp}
	D(\omega)=\left[\frac{n_e}{T_e}\left(1-\la\left(1-\frac{n\omega^*_e}{\ombar}\right)\left(1+\frac{n\omega_d}{\ombar}+\frac{n^2\omega^2_d}{\ombar^2}\right)\ra_t\right)+\sum_i\frac{n_iZ_i^2}{T_i}\left(1-\la\left(1-\frac{n\omega^*_i}{\ombar}\right)\left(1+\frac{n\omega_d}{\ombar}+\frac{n^2\omega^2_d}{\ombar^2}\right)\left(1-\frac{k_r^2\delta_i^2}{2}\right)\left(1-\frac{k_\theta^2\rho_i^2}{2}\right)\ra_t-\la\left(1-\frac{n\omega^*_i}{\ombar}\right)\left(1+\frac{n\omega_d}{\ombar}+\frac{k_\parallel{V}_\parallel}{\ombar}+\frac{k_\parallel^2V_\parallel^2}{\ombar^2}+\frac{2n\omega_dk_\parallel{V}_\parallel}{\ombar^2}+\frac{n^2\omega_d^2}{\ombar^2}\right)\left(1-\frac{k_\theta^2\rho_i^2}{2}-\frac{k_r^2\rho_i^2}{2}\right)\ra_p\right)\right]\phi(x)
\end{dmath}  
We have assumed a small wavenumber expansion in the Bessel functions. The Bessel functions are $J_0\left(k_\perp\rho_s\right)^2$ for the passing particles, and $J_0\left(k_\theta\rho_s\right)^2J_0\left(k_r\delta_s\right)^2$ for the trapped particles following both bounce and gyro-orbit averaging~\cite{sara05}. Since $\delta_e\ll\delta_i$ and $\rho_e\ll\rho_i$, we assume $J_0(k_r\delta_e)\approx1$ and $J_0(k_\perp\rho_e)\approx1$ for the electron terms.

$\la~\ra_t$ corresponds to the trapped particle velocity space integral, and $\la~\ra_p$ to the passing particle velocity space integral. The exponential term in the background Maxwellian is $e^{-\xi+M(2\vbar-M)}$.  We approximate with a Mach number expansion to first order, such that $e^{-\xi+M(2\vbar-M)}{\approx}e^{-\xi}\left(1+2M\vbar\right)$. In pitch angle coordinates, $\vbar=\epsilon_\parallel\sqrt{\xi\left(1-\lambda{b}\right)}$, where $b\equiv\frac{1+\bar{\epsilon}}{1+\bar{\epsilon}cos(\theta)}$ in the QuaLiKiz circular geometry, where ($\bar{\epsilon}\equiv\frac{r}{R}$. $\epsilon_\parallel$ notates a sum over -1 and 1, reflecting positive and negative velocities and leading to odd integrals of $\vbar$ leading to zero. For trapped particles, for any function of pitch angle and energy, the operation of the velocity space integration thus leads to:
\begin{dmath}
	\la{A(\xi,\kappa)}\ra_t{\equiv}\frac{2f_t}{\sqrt{\pi}}\int_{0}^{\infty}A\sqrt{\xi}e^{-\xi}d\xi\int_{0}^{1}K(\kappa)\kappa\left(1+2M\epsilon_\parallel\sqrt{\xi}\sqrt{1-\lambda{b}}\right){d\kappa}
\end{dmath}  
$\kappa$ is related to the pitch angle $\lambda$ by $\lambda\equiv1-2\bar{\epsilon}\kappa^2$. $K$ is the complete elliptic integral of the first kind, originating from bounce averaging in the small inverse aspect ratio expansion. We note that the poloidal dependence in $\vbar$ above, together with any poloidal dependence in A, must first be bounce averaged in the trapped particle integral. This is discussed below.

For passing particles:
\begin{dmath}
	\la{A(\xi,\kappa)}\ra_p{\equiv}\frac{2f_t}{\sqrt{\pi}}\int_{0}^{\infty}A\sqrt{\xi}e^{-\xi}d\xi\int_{0}^{\lambda_c}\frac{T(\lambda)}{4\pi}\left(1+2M\epsilon_\parallel\sqrt{\xi}\sqrt{1-\lambda{b}}\right){d\lambda}
\end{dmath}  
Where, in the small inverse aspect ratio expansion, the critical angle $\lambda_c=1-2\epbar$, and in the Jacobian: 
\begin{dmath} T(\lambda)=2\int_{0}^{\pi}\frac{1}{\sqrt{1-\lambda\left(1+2\epbar{sin^2(\theta/2)}\right)}}d\theta
\end{dmath}
Originating from the magnetic trapping perturbation to the transit frequency in the $dJ_2=dE/\omega_2$ term in the action-angle variable Jacobian, where $\omega_2$ is the passing particle angular transit frequency~\cite{bourphd}.  

The vertical drift frequency $n\omega_d$ in the QuaLiKiz shifted circle geometry is:
\begin{dmath} 
n\omega_d=-n\omega_{ds}\xi\left(2-\lambda{b}\right)\left(cos\theta+sin\theta\left(\shat\theta-\alpha\sin\theta\right)\right)
\end{dmath}
For the trapped particle integration, this expression is bounce averaged over the poloidal angle, becoming $n\omega_d=-n\omega_{ds}\xi{f(\kappa)}$, with:
\begin{dmath} 
	f(\kappa)=2\frac{E(\kappa)}{K(\kappa)}\left(1-\frac{4}{3}\kappa^2\alpha\right)-1+\left(4\shat+\frac{4}{3}\alpha\right)\left(\kappa^2-1+\frac{E(\kappa)}{K(\kappa)}\right)
\end{dmath}
Where $\alpha=q^2\sum_s\beta_s\left(\rlns+\rlts\right)$, and $K(\kappa)$ and $E(\kappa)$ are complete elliptic integrals of the first and second kinds, respectively. 

For the passing ions, we assume that the eigenfunction is strongly ballooned, and expand $\theta$ to second order, as well as invoking the small inverse aspect ratio expansion ($\epbar^2\to0$), leading to:
\begin{dmath} 
	n\omega_d=-n\omega_{ds}\xi\left(2-\lambda+\theta^2\left[\left(\shat-\alpha-\frac{1}{2}\right)\left(2-\lambda\right)-\frac{\lambda}{2}\epbar\right]\right)
\end{dmath}

The form of the eigenfunction $\phi(x)$ is assumed to be a shifted Gaussian:
\begin{dmath}
\phi(x)=\frac{\phi_0}{\left(\pi{\rm I\!R}(w^2)\right)^{0.25}}e^{-\frac{(x-x_0)^2}{2w^2}}
\end{dmath}
From the definition of the $k_r$ operator, $k_r^2=-\frac{\partial^2}{\partial{x^2}}$. Furthermore, with the strong ballooning approximation, the $\theta$ coordinate is related to the Fourier transform of the radial coordinate, leading to $\theta^2=k_r^2d^2$, where $d$ is the distance between rational surfaces. Assuming small shifts, $x_0{\ll}w^2$, the above operators act on $\phi(x)$ as follows:
\begin{dmath}
	\frac{\partial^2}{\partial{x^2}}\phi(x)=\left(\frac{x^2}{w^4}-\frac{2x_0x}{w^4}-\frac{1}{w^2}\right)\phi(x)
\end{dmath}
Finally, we also assume low flow shear, $\gamma_E\ll\omega$, leading to $\ombar^2\approx\omega^2-2{\omega}k_\theta\gamma_Ex$ and $\ombar^3\approx\omega^3-3\omega^2k_\theta\gamma_Ex$. 

With all the above definitions and assumptions, the approximated eigenfunction solution consists of inserting all the expressions into Eq.\ref{eq:disp}, carrying out the energy and pitch-angle integrations, and solving for $w$, $x_0$, and $\omega$. 

The energy integrals are carried out analytically, utilizing: $\la\xi\ra_\xi=\frac{3}{2}$, $\la\xi^2\ra_\xi=\frac{15}{4}$, $\la\xi\ra_\xi^3=\frac{105}{8}$, $\la\xi^4\ra_\xi=\frac{945}{16}$. The pitch-angle integrations are carried out numerically. 

The dispersion relation can be written in the form:
\begin{dmath}
	\label{eq:disp2}
	D(\omega)=\left(-\omega^3+3\omega^2k_\theta\gamma_Ex\right)\frac{n_e}{T_e}+F_{Te}\frac{n_e}{T_e}+\sum_i\left(-\omega^3+3\omega^2k_\theta\gamma_Ex\right)\frac{n_iZ_i^2}{T_i}+\sum_iF_{Ti}\frac{n_iZ_i^2}{T_i}+\sum_iF_{Pi}\frac{n_iZ_i^2}{T_i}=0
\end{dmath}
Where $F_{Te}$, $F_{Ti}$, $F_{Pi}$ contain all terms associated with the velocity space integrations for trapped electrons, trapped ions, and passing ions respectively. 

We summarize the ordering. Taking $\epbar$ as the small parameter, we assume:
\begin{eqnarray*}
	\mathcal{O}(k_\theta{x})\sim\mathcal{O}(d/w)\sim\mathcal{O}(\omega^*/\omega)&\sim&\mathcal{O}(1) \\
	\mathcal{O}(x/w)\sim\mathcal{O}(\delta_i/w)&\sim&\mathcal{O}(\epbar^{0.25}) \\
	\mathcal{O}(k_\theta\rho_i)\sim\mathcal{O}(\rho_i/w)\sim\mathcal{O}\left(max\left(\frac{L_T}{L_u},\frac{L_n}{L_u}\right)\right)&\sim&\mathcal{O}(\epbar^{0.5}) \\
	\mathcal{O}(\omega_d/\omega)\sim\mathcal{O}(k_\parallel{V_\parallel}/\omega)\sim\mathcal{O}(x_0/w)&\sim&\mathcal{O}(\epbar^{0.75}) \\ \mathcal{O}(M_i)\sim\mathcal{O}(\gamma_{E}/\omega)&\sim&\mathcal{O}(\epbar) \\ \mathcal{O}(M_e/M_i)\sim\mathcal{O}(\delta_e/w)\sim\mathcal{O}(\rho_e/w)&\sim&\mathcal{O}(\epbar^2)
\end{eqnarray*} 
We discard all terms $\mathcal{O}(\epbar^2)$ and higher. This leads to:

\begin{dmath}
	\label{eq:Te}
	F_{Te}=f_t\left(\omega^3-3k_\theta\gamma_E\omega^2x+\left(\omega^2-2k_\theta\gamma_E\omega{x}\right)W_{d1e0}-\frac{3}{2f_t}\left(\omega^2-2k_\theta\gamma_E\omega{x}\right)W^T_{V1e}-\frac{3}{2f_t}\left(\omega-k_\theta\gamma_Ex\right)W^T_{V1e}\left(W_{d1e0}+W_{d2e}\right)+\frac{3}{2f_t}W^{T2}_{V1e}\left(\omega+W_{d1e0}+W_{d2e}\right)\right)=F_{Te,0}+F_{Te,1}x
\end{dmath}

\begin{dmath}
	\label{eq:Ti}
	F_{Ti}=f_t\left(\dbar_i\left[\omega^3\rbar_i-3k_\theta\gamma_E\omega^2x+\omega^2\rbar_i\left(W_{d1i0}+W_{d1i1}\right)\right]-2k_\theta\gamma_E\omega{x}W_{d1i0}\dbar_i-\frac{3}{2f_t}\left(\omega^2\dhat_i-2k_\theta\gamma_E\omega{x}\right)W^T_{V1i}-\frac{3}{2f_t}\left(\omega\dhat_i-k_\theta\gamma_Ex\right)W^T_{V1i}\left(W_{d1i0}+W_{d2i}\right)-\frac{3}{2f_t}{\omega}W^T_{V1i}W_{d1i1}+\frac{3}{2f_t}W^{T2}_{V1i}\left(\omega+W_{d1i0}+W_{d2i}\right)+\frac{\delta^2_ix^2}{2w^4}\left(\omega^3+\omega^2\left(W_{d1i0}+W_{d1i1}-\frac{3}{2f_t}W^T_{V1i}\right)-\frac{3}{2f_t}\omega{W^T_{V1i}}W_{d1i0}\right)-\frac{\delta^2_ix_0x}{w^4}\left(\omega^3+\omega^2W_{d1i0}\right)\right)=F_{Ti,0}+F_{Ti,1}x+F_{Ti,2}x^2
\end{dmath}

\begin{dmath}
	\label{eq:Pi}
	F_{Pi}=f_p\left(\omega^3\rbar_i-3k_\theta\gamma_E\omega^2x-\frac{3}{2}\omega^2\left(W_{V1i}\rbar_i-W_{V2i0}\frac{d^2}{w^2}\left(\frac{x^2}{w^2}-\frac{2x_0x}{w^2}\right)+\frac{d^2}{w^2}\left(W_{V2i0}\rbar_i+W_{V2i1}\right)\right)+3k_\theta\gamma_E\omega{x}\left(W_{V1i}+\frac{d^2}{w^2}W_{V2i0}\right)+\frac{3}{2f_p}\kbar_i^2x^2\omega{V_1}+\left(\omega^2\rbar_i-2k_\theta\gamma_E\omega{x}\right)W_{d1i0}+\omega^2W_{d1i1}-\frac{3}{2}\omega\left(W_{d1i0}+W_{d2i}\right)\left(W_{V1i}\rbar_i-\frac{d^2}{w^2}W_{V2i0}\left(\frac{x^2}{w^2}-\frac{2x_0x}{w^2}\right)+\frac{d^2}{w^2}\left(W_{V2i0}\rbar_i+W_{V2i1}\right)\right)-\frac{3}{2}\omega{W_{d1i1}}\left(W_{V1i}+\frac{d^2}{w^2}W_{V2i0}\right)+\frac{3}{2}k_\theta\gamma_E{x}\left(W_{d1i0}+W_{d2i}\right)\left(W_{V1i}+\frac{d^2}{w^2}W_{V2i0}\right)+\frac{3}{2f_p}\kbar_i{x}{\omega}V_1\left(W_{d3i0}+W_{d3i1}\right)+\frac{3}{2f_p}\kbar_i^2x^2\left(W_{d1i0}+W_{d2i}\right)V_1+\frac{15}{4f_p}\omega\left(W^2_{V5i}+\frac{d^2}{w^2}W^2_{V6i0}\right)+\frac{15}{4f_p}\left(W_{d1i0}+2W_{d2i}\right)\left(W^2_{V5i}+\frac{d^2}{w^2}W^2_{V6i0}\right)-\frac{\rho_i^2}{2w^2}\left(\omega^3-\frac{3}{2}\omega^2\left(W_{V1i}+\frac{d^2}{w^2}W_{V2i0}-\frac{2}{3}W_{d1i0}\right)-\frac{3}{2}\omega\left(W_{d1i0}+W_{d2i}\right)\left(W_{V1i}+\frac{d^2}{w^2}W_{V2i0}\right)\right)+x^2\frac{\rho_i^2}{2w^4}\left(\omega^3+\omega^2W_{d1i0}\right)+\frac{2M}{f_p}\left[\frac{3}{2}\kbar_i{x}\omega^2V_{i1}+\frac{3}{2}\omega^2V_1W_{d3i0}+\frac{3}{2}\kbar_i{x}\omega\left(W_{d1i0}+W_{d2i}\right)V_1\right]\right)=F_{Pi,0}+F_{Pi,1}x+F_{Pi,2}x^2 
\end{dmath}

The definitions above are as follows. For the trapped species:
\begin{eqnarray*}
W^T_{V1s} &\equiv& n\omega_{ds}\la{f(\kappa)}\ra_\lambda  \\
W^{T2}_{V1s} &\equiv& n^2\omega_{ds}^2\la{f(\kappa)^2}\ra_\lambda  \\
V^T &\equiv& \la{2\epbar\frac{E\left(sin^{-1}\kappa,1/\kappa\right)}{\kappa{K(\kappa)}}}\ra_\lambda  \\
\end{eqnarray*} 
Where $E\left(sin^{-1}\kappa,1/\kappa\right)$ is an incomplete elliptic integral of the second kind, arising from the bounce averaged $V_\parallel^2$:
\begin{dmath}
V_\parallel^2(\lambda)=\frac{\oint\sqrt{1-\lambda{b}}}{\oint\frac{1}{\sqrt{1-\lambda{b}}}}=2\epbar\kappa\frac{E\left(sin^{-1}\kappa,1/\kappa\right)}{K(\kappa)}=2\epbar\left(\frac{E(\kappa)}{K(\kappa)}-(1-\kappa^2)\right)
\end{dmath}
Where we have utilized a convenient identity linking the incomplete elliptic integral of the second kind with the specific arguments we have obtained, to complete elliptic integrals of the first kind~\cite{wolf}.

For the passing species:
\begin{eqnarray*}
W_{V1s} &\equiv&  n\omega_{ds}\left(2-\frac{\la\lambda\ra_\lambda}{f_p}\right) \\
W_{V2s0} &\equiv&  n\omega_{ds}\left(\shat-\alpha-\frac{1}{2}\right)\left(2-\frac{\la\lambda\ra_\lambda}{f_p}\right) \\
W_{V2s1} &\equiv&  -n\omega_{ds}\epbar\frac{\la\lambda\ra_\lambda}{2f_p} \\
W_{V3s} &\equiv&  n\omega_{ds}\la\left(1-\lambda\right)\left(2-\lambda\right)\ra_\lambda \\
W_{V4s0} &\equiv&  n\omega_{ds}\la\left(\shat-\alpha-\frac{1}{2}\right)\left(2-\lambda\right)\left(1-\lambda\right)\ra_\lambda \\
W^2_{V5s} &\equiv& n^2\omega_{ds}^2\la\left(2-\lambda\right)^2\ra_\lambda \\
W^2_{V6s0} &\equiv&  n^2\omega_{ds}^2\la2\left(2-\lambda\right)\left(\shat-\alpha-\frac{1}{2}\right)\left(2-\lambda\right)\ra_\lambda \\
V_1 &\equiv&  1-\la\lambda\ra_\lambda \\
V_2 &\equiv&  \frac{-\epbar\la\lambda\ra_\lambda}{2} \\
\kbar_s &\equiv& {k_\theta}V_{Ts}\frac{\shat}{qR} \\
\rbar&\equiv&\left(1-\frac{k_\theta^2\rho_s^2}{2}\right) \\
\dbar&\equiv&\left(1-\frac{\delta_s^2}{2w^2}\right) \\
\dhat&\equiv&\left(1-\frac{\delta_s^2}{2w^2}-\frac{k_\theta^2\rho_s^2}{2}\right) \\
\end{eqnarray*}
Where $f_p$ is the passing particle fraction. Finally, common to both:
\begin{eqnarray*}
W_{d1s0} &\equiv& n\omega_{ds}\rlns\\
W_{d1s1} &\equiv& -2n\omega_{ds}M\rlus\\
W_{d2s} &\equiv& n\omega_{ds}\rlts\\
W_{d3s0} &\equiv& 2n\omega_{ds}\rlus\\
W_{d3s1} &\equiv& -2n\omega_{ds}M\rlts
\end{eqnarray*}
The $V$ expressions are geometric terms related to integrations over even powers of $\vbar$. The $W_V$ term arise from integrations including the vertical drift frequency. The $W_d$ terms arise from integrations over the diamagnetic frequencies. The subscript system is convenient due to the ordering of the various terms.

Following insertion of Eqs.~\ref{eq:Te},~\ref{eq:Ti},\ref{eq:Pi} into Eq.~\ref{eq:disp2}, the final step is to separate the terms proportional to $x^0$ (i.e., $1$), $x$, and $x^2$, such that:
\begin{equation}
	\label{eq:disp3}
	D(\omega)=D_0(\omega)+D_1(\omega)x+D_2(\omega)x^2=0
\end{equation}
With:
\begin{dmath}
	D_0(\omega)=-\omega^3\frac{n_e}{T_e}(1+Z_{eff}\frac{T_e}{T_i})+F_{Te,0}(\omega)\frac{n_e}{T_e}+\sum_i(F_{Ti,0}(\omega)+F_{Pi,0}(\omega))\frac{n_iZ_i^2}{T_e}
\end{dmath}
\begin{dmath}
	\label{eq:shift}
	D_1(\omega)=3\omega^2k_\theta\gamma_E\frac{n_e}{T_e}(1+Z_{eff}\frac{T_e}{T_i})+F_{Te,1}(\omega)\frac{n_e}{T_e}+\sum_i(F_{Ti,1}(\omega)+F_{Pi,1}(\omega))\frac{n_iZ_i^2}{T_i}
\end{dmath}
\begin{dmath}
		\label{eq:width}
	D_2(\omega)=\sum_i(F_{Ti,2}(\omega)+F_{Pi,2}(\omega))\frac{n_iZ_i^2}{T_i}
\end{dmath}

We have assumed all ion temperatures to be equal, allowing the emergence of $Z_{eff}$ in the ion summation.

Following Ref.~\cite{garb02}, we solve Eq.\ref{eq:disp3} in a perturbative manner. The eigenvalue $\omega_0$ is first solved from a modified form of $D_0(\omega)=0$, related to the local toroidal ITG dispersion relation at $\theta=0$, neglecting rotation, and setting $k_r=0$. Following multiplication by $T_e/n_e$ for convenience, this results in the following 3rd degree polynomial. 

\begin{dmath}
\label{eq:eigen}
C_3\omega^3+C_2\omega^2+C_1\omega+C_0=0	
\end{dmath}
Defining $\alpha_i{\equiv}Z_i^2\frac{n_i}{n_e}\frac{T_e}{T_i}$, we obtain:
%C_3\omega^3+C_2\omega^2+C_1\omega+C_0=0	
%C_2=f_tW_{d1e0}-\frac{3}{2}W^T_{V1e}+\sum_i\alpha_i\left(f_tW_{d1i0}-\frac{3}{2}W^T_{V1i}-\frac{3}{2}f_pW_{V1i}\rbar_i+f_p\rbar_iW_{d1i0}\right)
%C_1=-\frac{3}{2}W^T_{V1e}\left(W_{d1e0}+W_{d2e}\right)+\frac{3}{2}W^{T2}_{V1e}+\sum_i\alpha_i\left(-\frac{3}{2}W^T_{V1i}\left(W_{d1i0}+W_{d2i}\right)+\frac{3}{2}W^{T2}_{V1i}-\frac{3}{2}f_p\left(W_{d1i0}+W_{d2i}\right)W_{V1i}+\frac{15}{4}W^2_{V5i}\right)

\begin{dmath}
C_3=-1+f_t+\sum_i\alpha_i\left(-1+\rbar_i\right)
\end{dmath}

\begin{dmath}
C_2=f_tW_{d1e0}-\frac{3}{2}W^T_{V1e}+\sum_i\alpha_i\rbar_i\left(f_tW_{d1i0}-\frac{3}{2}W^T_{V1i}-\frac{3}{2}f_pW_{V1i}+f_pW_{d1i0}\right)
\end{dmath}
\begin{dmath}
C_1=-\frac{3}{2}W^T_{V1e}\left(W_{d1e0}+W_{d2e}\right)+\frac{3}{2}W^{T2}_{V1e}+\sum_i\alpha_i\left(-\frac{3}{2}W^T_{V1i}\rbar_i\left(W_{d1i0}+W_{d2i}\right)+\frac{3}{2}W^{T2}_{V1i}-\frac{3}{2}f_p\rbar_i\left(W_{d1i0}+W_{d2i}\right)W_{V1i}+\frac{15}{4}W^2_{V5i}\right)
\end{dmath}
\begin{dmath}
C_0=\frac{3}{2}W^{T2}_{V1e}\left(W_{d1e0}+W_{d2e}\right)+\sum_i\alpha_i\left(\frac{3}{2}W^{T2}_{V1i}\left(W_{d1i0}+W_{d2i}\right)+\frac{15}{4}\left(W_{d1i0}+2W_{d2i}\right)W^2_{V5i}\right)
\end{dmath}

Equations~\ref{eq:width} and ~\ref{eq:shift} then provide the solutions for the eigenfunction width and shift respectively, following substitution of the eigenvalue solution $\omega_0$ from equation~\ref{eq:eigen}. Here we maintain the rotation terms and finite $\theta$. This leads to the following expression for the eigenfunction width $w$. 
\begin{dmath}
\label{eq:x2}
w=\left(-\frac{\sum_i\alpha_i\left(\omega^3A_{3i}+\omega^2A_{2i}+\omega{A_{1i}}\right) }{\sum_i\alpha_i\kbar^2_i\left(\omega{B_{1i}}+B_{0i}\right)}\right)^{0.25}
\end{dmath}
with
\begin{dmath}
A_{3i}=f_t\frac{\delta^2_i}{2}+f_p\frac{\rho_i^2}{2}
\end{dmath}
\begin{dmath}
A_{2i}=f_t\frac{\delta_i^2}{2}\left(W_{d1i0}+W_{d1i1}-\frac{3}{2f_t}W^T_{V1i}\right)+f_p\frac{\rho_i^2}{2}W_{d1i0}+f_p\frac{3}{2}d^2W_{V2i0}
\end{dmath}
\begin{dmath}
A_{1i}=-\frac{\delta_i^2}{2}\frac{3}{2}W^T_{V1i}\left(W_{d1i0}+W_{d2i}\right)+\frac{3}{2}f_pd^2\left(W_{d1i0}+W_{d2i}\right)W_{V2i0}
\end{dmath}
\begin{dmath}
B_{1i}=\frac{3}{2}V_1
\end{dmath}
\begin{dmath}
B_{0i}=\frac{3}{2}V_1\left(W_{d1i0}+W_{d2i}\right)
\end{dmath}

The eigenfunction shift is given by: 
\begin{dmath}
\label{eq:x}
x_0=-\frac{\gamma_E\left(U_{1e}+\sum_i\alpha_iU_{1i}\right)+\sum_i\alpha_iU_{2i}}{\sum_i\alpha_iD_i}
\end{dmath}
with:
\begin{dmath}
U_{1e}=3\omega^2k_\theta\left(1-f_t\right)-2k_\theta\omega{f_t}W_{d1e0}+3k_\theta\omega{W^T_{V1e}}+\frac{3}{2}k_\theta{W^T_{V1e}}\left(W_{d1e0}+W_{d2e}\right)
\end{dmath}
\begin{dmath}
	U_{1i}=3\omega^2k_\theta\left(1-f_t\dbar_i\right)-2k_\theta\omega{f_t}\left(W_{d1i0}\dbar_i+W_{d1i1}\right)+3k_\theta\omega{W^T_{V1i}}+\frac{3}{2}k_\theta{W^T_{V1i}}\left(W_{d1i0}+W_{d2i}\right)-3k_\theta{f_p}\omega^2+3k_\theta{f_p}\omega\left(W_{V1i}+\frac{d^2}{w^2}W_{V2i0}\right)-2k_\theta\omega{f_p}\left(W_{d1i0}+W_{d1i1}\right)+\frac{3}{2}k_\theta{f_p}\left(W_{d1i0}+W_{d2i}\right)\left(W_{V1i}+\frac{d^2}{w^2}W_{V2i0}\right)
\end{dmath}
\begin{dmath}
U_{2i}=\frac{3}{2}\kbar_iV_1\left(\omega\left(W_{d3i0}+W_{d3i1}\right)+2M\left(\omega^2+\omega\left(W_{d1i0}+W_{d2i}\right)\right)\right)
\end{dmath}
\begin{dmath}
D_i=-\frac{\delta_i}{w^4}f_t\left(\omega^3+\omega^2W_{d1i0}\right)-3f_p\omega^2W_{V2i0}\frac{d^2}{w^4}-3f_p\omega\frac{d^2}{w^4}W_{V2i0}\left(W_{d1i0}+W_{d2i}\right)
\end{dmath}

There are two tuning factors in this process. Firstly, the eigenfunction width, before insertion into equation ~\ref{eq:x}, is multiplied by a factor $\left(\frac{0.1}{k_\theta\rho_s}\right)^{0.25}$. This factor was tuned by comparison to full gyrokinetic predictions at the GA-standard case. Then, following this, the width is normalized by the factor $max(1,0.66d/w)$, where $d$ is the distance between neighbouring rational surfaces. This is to reduce the occasional emergence of non-physical narrow width solutions, which can significantly slow down the full kinetic dispersion relation solution procedure. 
 
\subsection{Electron scales}
For the electron scale fluid eigenfunction solution, we assume adiabatic ions, and now include passing electrons. Electron scales are defined in QuaLiKiz as $k_\theta\rho_s>2$. The ordering is similar, apart from $\mathcal{O}(\delta_e/w)\sim\mathcal{O}(\epbar^{0.25})$, $\mathcal{O}(k_\theta\rho_e)\sim\mathcal{O}(\epbar^{0.5})$, $\mathcal{O}(\omega_d/\omega)\sim\mathcal{O}(\epbar)$, and $\mathcal{O}(\gamma_E/\omega)\sim\mathcal{O}(\epbar^2)$. The dispersion relation is now simpler:
\begin{dmath}
D(\omega)=\left[\la\left(1-\frac{n\omega^*_e}{\omega}\right)\left(1+\frac{n\omega_d}{\omega}\right)\dbar_e\rbar_e\ra_t+\la\left(1-\frac{n\omega^*_e}{\omega}\right)\left(1+\frac{n\omega_d}{\omega}+\frac{k_\parallel{V}_\parallel}{\omega}+\frac{k^2_\parallel{V^2_\parallel}}{\omega^2}+\frac{2n\omega_d{k_\parallel}V_\parallel}{\omega^2}\right)\rbar_e\ra_p-1-\frac{T_e}{T_i}Z_{eff}\right]=0
\end{dmath}
Which can be written as:
\begin{equation}
\label{eq:dispETG}
D(\omega)=F^{ETG}_T+F^{ETG}_P-\omega^3\left(1+\tau_e\right)=0
\end{equation}
With $\tau_e\equiv\frac{T_e}{T_i}Z_{eff}$
The expressions for $F_{Te}$ and $F_{Pe}$ are also relatively simpler, due to the ordering out of any rotation related terms. 
\begin{dmath}	
F^{ETG}_T=f_t\left[\left(\omega^3\rbar_e+\omega^2\rbar_eW_{d1e0}-\frac{3}{2f_t}\omega^2W^T_{V1e}-\frac{3}{2f_t}\omega\left(W_{d1e0}+W_{d2e}\right)W^T_{V1e}\right)\dbar_e+\frac{\delta^2_ex^2}{2w^4}\left(\omega^3+\omega^2W_{d1e0}\right)\right]
\end{dmath}
\begin{dmath}	
	F^{ETG}_P=f_p\left[\omega^3\rbar_e-\frac{3}{2}\omega^2\left(W_{V1e}-W_{V2e0}\frac{d^2}{w^2}\left(\frac{x^2}{w^2}-1\right)\right)+\frac{3}{2f_p}\kbar^2_e{x^2}\omega{V_1}+\omega^2W_{d1e0}\rbar_e-\frac{3}{2}\omega\left(W_{d1e0}+W_{d2e}\right)\left(W_{V1e}-W_{V2e0}\frac{d^2}{w^2}\left(\frac{x^2}{w^2}-1\right)\right)+\frac{3}{2f_p}\kbar^2_e{x^2}\left(W_{d1e0}+W_{d2e}\right)V_1+\frac{\rho_e^2}{2w^2}\left(\frac{x^2}{w^2}-1\right)\left(\omega^3+\omega^2W_{d1e0}\right)\right]
\end{dmath}
The rest of the ETG solution proceeds as for the ion scales. Equation \ref{eq:dispETG} is solved by separating the terms proportional to $x^0$ and $x^2$, and setting each resultant equation to zero. In the perturbative scheme, the $x^0$ polynomial equation is solved at $\theta=0$ and setting $k_r=0$, providing the toroidal ETG eigenvalue. The full equation with terms proportional to $x^2$ then provides the eigenfunction width, when substituting the eigenvalue from the polynomial solution. The coefficients in the polynomial equation are as follows:
\begin{dmath}
	\label{eq:1ETG}
	C_2\omega^2+C_1\omega+C_0=0	
\end{dmath}
with:
\begin{dmath}
	C_2=-1-\tau_e+\rbar_e
\end{dmath}
\begin{dmath}
	C_1=W_{d1e0}f_t\rbar_e-\frac{3}{2}W^T_{V1e}-\frac{3}{2}f_pW_{V1e}+W_{d1e0}f_p\rbar_e
\end{dmath}
\begin{dmath}
	C_0=-\frac{3}{2}\left(W_{d1e0}+W_{d2e}\right)W^T_{V1e}-\frac{3}{2}f_p\left(W_{d1e0}+W_{d2e}\right)W_{V1e}
\end{dmath}
For the width equation, we obtain:  
\begin{dmath}
\label{eq:x2}
w=\left(-\frac{\omega^3A_3+\omega^2A_2+\omega{A_1}}{\kbar^2_e\left(\omega{B_1}+B_0\right)}\right)^{0.25}
\end{dmath}
with:
\begin{dmath}
	A_3=f_t\frac{\delta^2_e}{2}+f_p\frac{\rho^2_e}{2}
\end{dmath}
\begin{dmath}
	A_2=f_t\frac{\delta_e^2}{2}W_{d1e0}+f_p\frac{\rho_e^2}{2}W_{d1e0}+\frac{3}{2}f_pd^2W_{V2e0}
\end{dmath}
\begin{dmath}
	A_1=\frac{3}{2}f_pd^2\left(W_{d1e0}+W_{d2e}\right)W_{V2e0}
\end{dmath}
\begin{dmath}
	B_1=\frac{3}{2}V_1
\end{dmath}
\begin{dmath}
	B_0=\frac{3}{2}V_1\left(W_{d1e0}+W_{d2e}\right)
\end{dmath}

\section{Trace heavy impurity transport transport coefficients with asymmetries}
\label{app:asym}
The impact of asymmetries on the transport coefficients is through the weighted flux surface average of the modified $n$ and $\frac{dn}{dr}$ in the quasilinear transport integrals. The averages are weighted by the electrostatic potential eigenfunction. 

We closely follow the procedure outlined in Ref.~\cite{angi12}, with two generalizations. We include temperature anisotropies, and also build the transport coefficients from a spectrum of quasilinear weights. 

The modified density is:
\begin{dmath}	
n_j(r,\theta)=n_{r,LFS}(r)P_j(r,\theta)exp\left(-\frac{\epsilon_j(r,\theta)}{T_{\parallel{j}}(r)}\right)
\end{dmath}
with
\begin{dmath}	
\epsilon_j=Z_je\Phi(r,\theta)-\frac{m_j\Omega^2(r)}{2}\left(R^2(\theta,r)-R^2_{LFS}(r)\right)
\end{dmath}
and
\begin{dmath}	
P(r,\theta)\equiv\frac{T_{\perp{j}}(r,\theta)}{T_{\perp{j},LFS}(r)}=\left[\frac{T_{\perp{j},LFS}(r)}{T_{\parallel{j},LFS}(r)}+\left(1-\frac{T_{\perp{j},LFS}(r)}{T_{\parallel{j},LFS}(r)}\right)\frac{B_{LFS}(r)}{B(r,\theta)}\right]^{-1}
\end{dmath}
where $\Phi(r,\theta)$ is solved by imposing the quasineutrality constraint at each poloidal location. 

We define the following weighted flux surface averaging:
\begin{dmath}
\la{A(\theta)}\ra\equiv\frac{{\oint}A(\theta)\left(1+{\epbar}cos\theta\right)e^{-\frac{1}{2}\frac{w^2}{d^2}\theta^2}d\theta}{\oint\left(1+\hat{\epsilon}cos\theta\right)e^{-\frac{1}{2}\frac{w^2}{d^2}\theta^2}d\theta}
\end{dmath}
$d$ is the distance between neighbouring rational flux surfaces, and $w$ is the eigenfunction width as calculated by QuaLiKiz. 
The following expressions are then convenient. They are similar to those in Ref.~\cite{angi14}, and are related to various contributions to $n$ and $\frac{dn}{dr}$:
\begin{eqnarray*}
	e_0 &\equiv&  \la{Pe^{-\epsilon/T}}\ra \\
	e_1 &\equiv&  \la{\frac{Ze}{T}{\Phi}Pe^{-\epsilon/T}}\ra \\
	e_2 &\equiv&  R_0\la{\frac{Ze}{T}{\Phi}Pe^{-\epsilon/T}}\ra \\
	e_3 &\equiv&  \frac{1}{R_0^2}\la{\left(R^2(\theta)-R^2_{LFS}\right)Pe^{-\epsilon/T}}\ra \\
	e_4 &\equiv&  \frac{1}{R_0}\la{\left(R(\theta)\frac{R(\theta)}{dr}-R_{LFS}\frac{dR_{LFS}}{dr}\right)Pe^{-\epsilon/T}}\ra \\
	e_6 &\equiv&  R_0\la{\frac{dP}{dr}e^{-\epsilon/T}}\ra \\
	e_7 &\equiv&  \la{\frac{Ze}{T}Pe^{-\epsilon/T}\frac{d\Phi}{d\theta}\theta\frac{\hat{s}}{\epbar}}\ra \\
	e_8 &\equiv&  \la{Pe^{-\epsilon/T}\left(1+{\epbar}cos\theta\right)\left(cos\theta+\shat\theta{sin\theta}\right)}\ra \\
	e_9 &\equiv& \frac{2}{R_0}\la{-R_{LFS}\frac{dR_{LFS}}{dr}Pe^{-\epsilon/T}}\ra 
\end{eqnarray*}
Where $R_0$ is the normalizing major radius.

The particle transport coefficients (for a given species) in the QuaLiKiz output, are then defined as follows, with respect to the input low-field-side densities and density gradient:

\begin{dmath}
\Gamma=-\bar{D}\frac{dn_{LFS}}{dr}+\left(\bar{V_T}+\bar{V_C}+V_{TCF}+V_{UCF}+V_{PCF}\right)n_{LFS} 
\end{dmath}
Where $\bar{D}$, $\bar{V_T}$, and $\bar{V_C}$ are the poloidal asymmetry modified diffusivity, thermodiffusive pinch, and compressional pinch respectively.
\begin{eqnarray*}
\bar{D} &=&  \sum_kD_k{e_{0k}} \\
\bar{V_T} &=&  \sum_kV_{Tk}{e_{0k}} \\
\bar{V_C} &=&  \sum_kV_{Ck}{e_{0k}} 
\end{eqnarray*}
$D_k$, $V_{Tk}$, and $V_{Ck}$ are the original wavenumber dependent diffusivity and pinch terms, as calculated from the quasilinear integrals without poloidal asymmetries, and already include all nonlinear saturation rule spectral weights and normalizations. The summation over $k$ is necessary due to the wavenumber dependent eigenfunction weighting in $e_0$.

Borrowing from the terminology in Ref.~\cite{angi12}, $V_{TCF}$, $V_{UCF}$, and $V_{PCF}$ can be interpreted as additional thermodiffusive pinch, rotodiffusive pinch, and pure pinch terms, arising due to the poloidal asymmetries. Physically, they arise from a modification of the effective diffusion coefficient due to poloidal asymmetry in the density gradient. The terms are the following:
\begin{eqnarray*}
	V_{TCF} &=&  \sum_k\frac{D_k}{R_0}\left(\frac{R}{L_{T}}\left(e_{1k}-\frac{m\Omega^2R_0^2}{2T}e_{3k}\right)\right) \\
	V_{UCF} &=&  \sum_k2\frac{D_k}{R_0}M\frac{dM}{dr}e_{3k} \\
	V_{PCF} &=&  \sum_k\frac{D_k}{R_0}\left(e_{2k}-e_{7k}-2M^2\left(e_{6k}+e_{8k}+\frac{e_{9k}}{2}\right)\right) \\
\end{eqnarray*}
The main differences between this formulation and Ref.~\cite{angi12} are: the summation over $k$ for the total quasilinear flux; the inclusion of the $P$ and $e_6$ temperature anisotropy terms.

The QuaLiKiz outputted transport coefficients are all with respect to $n_{LFS}$ and $\frac{dn_{LFS}}{dr}$. In a transport code, such as JETTO, where the evolved densities are rather the flux-surface-average $FSA$ quantities, then the transport coefficients must be transformed to be respective to the $FSA$ quantities. 

\end{appendix}
\end{document}